\documentclass[twocolumn,english,prb,epsfig,rotate,showpacs,aps]{revtex4-1}

\usepackage[T1]{fontenc}
\usepackage[latin9]{inputenc}
\usepackage{color}
\usepackage{verbatim}
\usepackage{amsmath}
\usepackage{amssymb}
\usepackage{graphicx}
\usepackage{esint}
\usepackage{microtype}
\usepackage{hyperref}
\usepackage{lmodern}
\makeatletter
\usepackage{epsfig}
\usepackage{amsmath}
\usepackage{graphicx}
\usepackage{dcolumn}
\usepackage{bm}
\usepackage{mathtools}
\usepackage{amsfonts,amssymb}
\usepackage[T1]{fontenc}
\usepackage[latin9]{inputenc}
\usepackage{amsmath}
\usepackage{graphicx}
\usepackage{amssymb}
\usepackage{esint}
\usepackage{babel}
\usepackage{microtype}
\usepackage{xcolor}
\usepackage{ulem}
\usepackage{natbib}
\newcommand{\bk}{\mathbf{k}}
\newcommand{\bq}{\mathbf{q}}
\newcommand{\br}{\mathbf{r}}

\newcommand{\bK}{\mathbf{K}}
\newcommand{\brho}{\boldsymbol{\rho}}

\newcommand{\red}{\color{black}}
\newcommand{\blue}{\color{black}}

\usepackage{titlesec}
\titlespacing*{\section}{0pt}{2ex plus 1ex minus .2ex}{2ex plus .2ex}

\begin{document}

\setlength{\baselineskip}{11pt}
\title{Chiral kinematic theory and converse vortical effects}
\author{Kai Chen }
\affiliation{Department of Physics and Texas Center for Superconductivity, University of Houston, Houston, TX 77204}
\author{Swadeepan Nanda}
\affiliation{Department of Physics and Texas Center for Superconductivity, University of Houston, Houston, TX 77204}
\author{Pavan Hosur}
\affiliation{Department of Physics and Texas Center for Superconductivity, University of Houston, Houston, TX 77204}
\date{\today}
%\email{kchen3@graduatecenter.cuny.edu }

%\maketitle
\begin{abstract}
Response theories in condensed matter typically describe the response of an electron fluid to external electromagnetic fields, while perturbations on neutral particles are often designed to mimic such fields. Here, we study the response of fermions to a space-time-dependent velocity field, thereby sidestepping the issue of gauge charge. First, we use a semiclassical chiral kinematic theory to obtain the local density of current and extract the orbital magnetization. The theory immediately predicts a "converse vortical effect," defined as an orbital magnetization driven by linear velocity. It receives contributions from magnetic moments on the Fermi surface and the Berry curvature of the occupied bands. Then, transcending semiclassics via a complementary Kubo formalism reveals that the uniform limit of a clean system receives only the Berry curvature contribution while other limits sense the Fermi surface magnetic moments too. We propose CoSi as a candidate material and suggest magnetometry of a sample under a thermal gradient to detect the effect. Overall, our study sheds light on the effects of a space-time-dependent velocity field on electron fluids and paves the way for exploring quantum materials using new probes and perturbations.
\end{abstract}
\maketitle

\section{Introduction}

Response theories, a fundamental framework in physics, explore how physical systems dynamically respond to external perturbations. In the context of quantum materials, they describe a myriad of properties ranging from conventional ones such as longitudinal conductivity and magnetization, to topological ones such as the quantized Hall conductivity of two-dimensional (2D) insulators and the half-quantum Hall effect on the surface of 3D topological insulators \cite{hasan2010colloquium,qi2011topological}. As most responses involve the constituent electrons responding to external electromagnetic fields, response theories provide a bridge from microscopic quantum phenomena to macroscopic material properties and facilitate the design of novel functional materials tunable by these fields.

%In condensed matter physics, electrons play a pivotal role in responding to external perturbations, leading to intriguing transport phenomena. While the behavior of free particles under such perturbations can be intuitively understood, the inclusion of lattice structures adds complexity to the picture. An exemplary case is the modification of group velocities due to the anomalous velocity arising from the Berry curvature of Bloch states \cite{sundaram1999wave,chang1995berry,chang1996berry,xiao2010berry, dong2018geometrodynamics}. 

A striking family of responses with deep topological roots are the so-called chiral responses. Chirality refers to an intrinsic handedness of the system and is non-zero only in systems that break all improper symmetries including parity, such as an isolated Weyl fermion. Thus, chiral responses were initially explored in various contexts in fundamental physics ranging from left-handed neutrinos \cite{yanagida1980horizontal,gelmini1981left} and parity violation \cite{lee1956question,wu1957experimental} in the Standard Model to the fluid dynamics of rotating blackholes \cite{vilenkin1979macroscopic,yamamoto2016chiral,prokhorov2020chiral,flachi2018chiral} and axion models of dark matter \cite{alexander2018chiral,di2021breakdown,chadha2022axion}. Since the discovery of Weyl semimetals (WSMs), whose band structure contains intersections around which the excitations resemble relativistic Weyl fermions, \cite{armitage2018weyl,hosur2013recent,wan2011topological,burkov2011weyl,lv2015experimental,soluyanov2015type,weng2015weyl,xu2015discovery}, interest has mushroomed in chiral responses in condensed matter \cite{zyuzin2012topological,li2016chiral,ma2015chiral,huang2017topological,ong2021experimental,jia2016weyl}. Most chiral responses can be traced to topological chiral anomalies, defined as the breakdown of classical conservation laws upon quantization of chiral fermion \cite{adler1969axial,zumino1984chiral,alekseev1998universality,jia2016weyl,zhang2016signatures,mueller2018chiral,rylands2021chiral}. The anomalies, too, were first explored in high-energy physics, but have found remarkable applications in topological condensed matter, particularly in Weyl and Dirac semimetals, manifesting as exotic transport phenomena \cite{gorbar2014chiral,potter2014quantum,xiong2015evidence,grushin2016inhomogeneous,burkov2016topological,liang2017anomalous}.

An important and fundamental chiral response is the chiral vortical effect (CVE), defined as an axial current driven by rotation in a chiral fluid \cite{yin,chen2014lorentz}. Since the advent of WSMs, it has been derived for general chiral band structures \cite{bacsar2014triangle,landsteiner2014anomalous,shitade2020chiral}, has been extended to arbitrary spacetime-dependent rotation, and motivated a related effect dubbed the gyrotropic vortical effect (GVE) \cite{nanda2023vortical}. Moreover, the generalizations have revealed the origins of the effect in the Berry curvature $\boldsymbol{\Omega}_{n}(\mathbf{k})$ and orbital magnetic moment $\mathbf{m}_{n}^\text{orb}(\mathbf{k})$ of the relevant bands.

A powerful framework that captures chiral responses including the topological nature of the chiral anomaly is the semiclassical chiral kinetic theory \cite{son2012berry,son2013chiral,son2013kinetic,yin,burkov2015chiral,liu2019chiral,kharzeev2016chiral,chen2016nonlinear,hidaka2018nonlinear,mameda2023nonlinear}.
% Here, the topological content is encoded in the enhancement of the phase space measure in the $n$-th band by $1+\mathbf{B}\cdot\boldsymbol{\Omega}_n(\mathbf{k})$, where $\mathbf{B}$ is the magnetic field and $\boldsymbol{\Omega}_n(\mathbf{k})$ is the Berry curvature of the band at momentum $\mathbf{k}$.
Moreover, using analogies between electromagnetic and fictitious non-inertial fields, such as the similarity between the classical Lorentz and Coriolis forces, chiral kinetic theory can also encompass certain responses of Weyl fermions to space and time-dependent velocity fields $\mathbf{v}(\mathbf{r},t)$ including the CVE \cite{nanda2023vortical,yin,dayi2018quantum,dayi2021quantum,shitade2020chiral,toshio2020anomalous}. Such kinematic responses -- chiral and otherwise -- are routinely used to simulate gauge fields for neutral ultracold atoms \cite{cooper2008rapidly,lin2009synthetic,fetter2009rotating,goldman2014light}. They are arguably more fundamental than electromagnetic responses as they do not rely on a well-defined conserved charge and exist, for instance, even for superconducting quasiparticles whose charge is ill-defined. However, while the analogies are established for non-relativistic and relativistic free particles in vacuum, they are unknown for electrons in general band structures. Thus, a general description of kinematic responses independently of electromagnetic analogies is highly desirable.

%The semiclassical kinetic theory for Dirac and Weyl particles with electromagnetic fields and global rotation is studied in Ref. \cite{dayi2017semiclassical}, and an effective curved-space Weyl theory is presented in Ref. \cite{liang2019curved}.

%The outline of this paper is as follows. 
In this paper, we study the linear response of electrons in general band structures to a space-time dependent velocity field, $\mathbf{v}(\mathbf{r},t)$. Interestingly, if the band structure is chiral, we show that the velocity field gives rise to an orbital magnetization. The effect is the opposite of the vortical effect in a sense that we specify, and we dub the phenomenon the converse vortical effect. 

We begin by introducing the converse vortical effect via an analogy with the vortical effect in Sec. \ref{sec:effects}. We then use a semiclassical chiral kinematic theory to derive this effect in Sec. \ref{sec:ckmt}. In particular, we calculate the orbital magnetization $\mathbf{M}^\text{orb} = \chi^\text{orb} \mathbf{\cdot v}$, where $\chi^\text{orb}$ denotes the susceptibility of orbital magnetization to the velocity field.
%We refer to this phenomenon as the converse vortical effect.
%\begin{equation}
%\mathbf{M}^\text{orb} = \chi \mathbf{v}
%\end{equation} 
%that we term the converse vortical effect. 
Then, in Sec. \ref{sec:Kubo}, we employ a complementary, quantum mechanical Kubo approach to compute the linear response function at general frequencies $\omega$ and momenta $\mathbf{q}$ of the velocity field in the presence of a phenomenological quasiparticle lifetime $\tau$. This approach shows that the uniform limit of a clean system, defined by $q=0$, $\omega\to0$ and $|\omega\tau|\gg1$, has a response purely governed by $\boldsymbol{\Omega}_n(\mathbf{k})$ of the occupied bands. In contrast, other orders of limits of $\omega\to0$, $q\to0$ and $\tau\to\infty$ also acquire contributions from $\mathbf{m}^\text{orb}_n(\mathbf{k})$ on the Fermi surface. We refer to the response in the uniform limit ($q\to0$ before $\omega\to0$) as the converse gyrotropic vortical effect (cGVE), and that in the static limit ($q\to0$ after $\omega\to0$) as the converse chiral vortical effect (cCVE).
In Sec. \ref{sec:Weyl}, we calculate the cGVE and cCVE in Weyl fermions, and close by proposing CoSi as a candidate material to observe both the cGVE and cCVE in Sec. \ref{expt}.

\section{Vortical and Converse Vortical Effects}\label{sec:effects}

The vortical and converse vortical effects can be heuristically likened to a bolt and nut analogy. When the head of a bolt is rotated faster, it generates more torque, which transforms into linear force, enabling the bolt to move faster inside the nut. This phenomenon of linear motion driven by circular resembles the vortical effect. Conversely, when a bolt has a higher linear speed inside the nut, its head gains faster circular rotation. This is the classical counterpart of the converse vortical effect. The converse vortical effect is a quantum phenomenon and, as we will see later, relies on the Berry curvature and magnetic moment of Bloch electrons.
%We emphasize the quantum nature of the converse vortical effect, distinct from its classical analogy with a bolt and nut. This effect links velocity to orbital magnetization, which relies on the Bloch wave function of electrons in the lattice background. Consequently, there may exist a fundamental connection between orbital magnetization and the Berry phase, a fundamental quantum phenomenon.

%The converse vortical effect is governed by two crucial factors: the orbital magnetic moment $\mathbf{m}^\text{orb}_n(\mathbf{k})$ on the Fermi surface and $\boldsymbol{\Omega}_n(\mathbf{k})$ of occupied bands, the latter stemming from the modified phase space measure. A similar separation of contributions occurs in modern theories of orbital magnetization based on semiclassical wave packet dynamics \cite{sundaram1999wave,xiao2006berry,ceresoli2006orbital,thonhauser2005orbital,xiao2021adiabatically,dong2018geometrodynamics} or quantum perturbation theory \cite{shi2007quantum} in electromagnetic fields. 
\begin{figure}[h]
%\centering
\includegraphics[width=0.95\columnwidth]{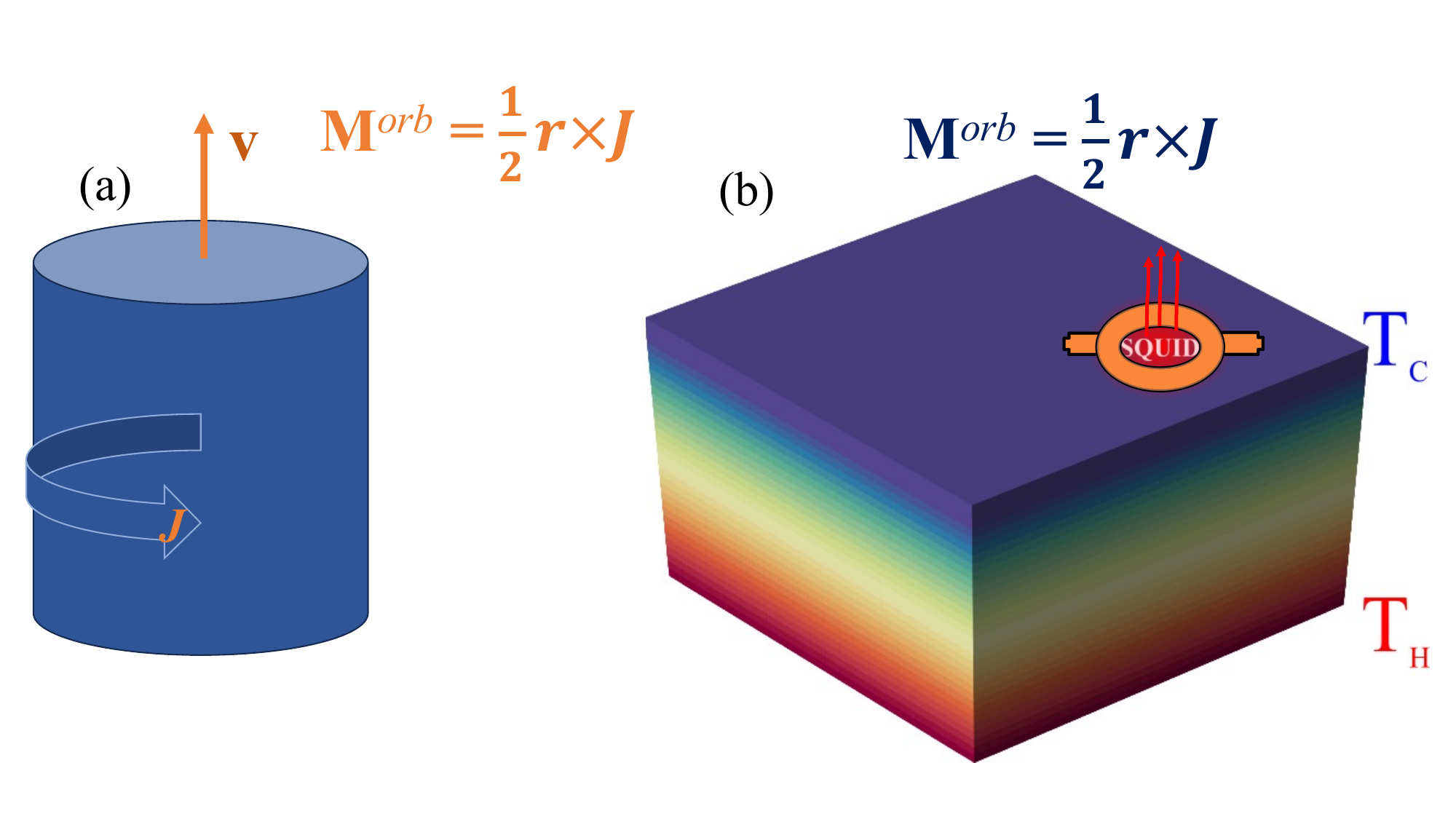}
\caption{ Color online. (a) Schematic depiction of the converse vortical effect. (b) Device geometry for observing the converse vortical effect: The temperature difference propels electrons, inducing their motion with velocity $\mathbf{v}$. The resulting magnetization can be observed aligned with the velocity field.}
\label{Fig.1}
%\end{center}
\end{figure}
To better understand the converse effects -- and their nomenclature -- in a broader context, let us recap other closely related effects. First, continuum Weyl fermions in a $\mathbf{B}$-field exhibit the chiral magnetic effect (CME), arising from the chiral anomaly and manifesting as a current parallel to $\mathbf{B}$ \cite{li2016chiral,hosur2013recent,armitage2018weyl}, unlike conventional charged particles that undergo circular motion in an orthogonal plane. In WSMs, the CME vanishes at equilibrium due to Bloch's theorem \cite{bohm1949note,yamamoto2015generalized} that forbids equilibrium current densities but persists in non-equilibrium steady states with unequal Fermi levels for left- and right-handed Weyl nodes. Reconciling the continuum and lattice manifestations of the CME involved considering non-zero $\mathbf{q}$ and $\omega$ responses. The original CME emerges in the static limit and relies on the existence of Weyl nodes while the uniform limit revealed a new effect, termed the gyrotropic magnetic effect (GME) \cite{ma2015chiral,zhong2016gyrotropic}, that corresponds to a current along a time-dependent $\mathbf{B}$-field and exists for general band structures. Analogous to the CME, the CVE corresponds to the static limit and represents another anomaly-induced transport phenomenon \cite{khaidukov2012chiral,chen2014lorentz,shitade2020chiral,toshio2020anomalous,kirilin2012chiral,avkhadiev2017chiral,abramchuk2018anatomy,gao2019chiral}, namely, the dissipationless axial current proportional to $\boldsymbol{\mathcal{V}}=\frac{1}{2}\boldsymbol{\nabla}\times\mathbf{v}$. Similarly, the gyrotropic vortical effect (GVE) was recently defined as the extension of the CVE to the uniform limit that crucially relies on the time-dependence of $\boldsymbol{\mathcal{V}}$ \cite{nanda2023vortical}. Both vortical effects exist at equilibrium for general band structures regardless of Weyl nodes. In short, the CME and GME are defined by $\mathbf{J}\propto\mathbf{B}$ in different limits whereas the CVE and GVE are given by $\mathbf{J}\propto\boldsymbol{\mathcal{V}}$ in these limits.
 
The GME inspires an effect dubbed the inverse GME, defined as magnetization proportional to the vector potential, $\mathbf{M} \propto \mathbf{A}$ with a response function that is the matrix inverse of that of the GME \cite{zhong2016gyrotropic}. The CME lacks an inverse response since a static $\mathbf{A}$ is a pure gauge field. The GME and its inverse are related by an interchange of conjugate variables, $\mathbf{J}\leftrightarrow\mathbf{A}$ and $\mathbf{B}\leftrightarrow\mathbf{M}$, followed by an interchange of the left- and right-hand sides. Physically, this means the field $\mathbf{A}$ conjugate to the GME response $\mathbf{J}$ drives the inverse GME, and vice versa. This pattern suggests analogous inverse kinematic responses where a linear momentum $\mathbf{K}$, conjugate to the linear current  $\mathbf{J}$, drives an angular momentum $\mathbf{L}$ that is conjugate to the angular velocity $\boldsymbol{\mathcal{V}}$. While such responses presumably exist, our focus is on a distinct class of effects: unlike the inverse effects, the converse effects correspond to an interchange of responding and driving fields without conjugation. Thus, we wish to compute an angular velocity $\boldsymbol{\mathcal{V}}$ driven by a linear velocity $\mathbf{v}$. However, we compute a slightly different quantity that also characterizes rotational motion, namely, $\mathbf{M}^\text{orb} \equiv \frac{1}{2}\mathbf{r}\times\mathbf{J}$ as a proxy for $\boldsymbol{\mathcal{V}}$, as $\mathbf{M}^\text{orb}$ is directly measurable in experiments and easier to compute than $\boldsymbol{\mathcal{V}}$. Note that in general, $\mathbf{L}$, not $\mathbf{M}^\text{orb}$, is conjugate to $\boldsymbol{\mathcal{V}}$ and the diamagnetic contribution to the current is $\boldsymbol\nabla\times\mathbf{M}^\text{orb}$. Only in simple cases such as a classical current loop, $\mathbf{L}\propto\mathbf{M}^\text{orb}$, and $\mathbf{M}^\text{orb}$ and $\mathbf{L}$ are used interchangeably upto overall constants. 

\section{Chiral kinematic theory}\label{sec:ckmt}

{\blue{ In this section, we first employ the semiclassical wave-packet dynamics method \cite{sundaram1999wave} to obtain the response function for the converse vortical effect. We will first derive the equations of motion (EM) for electrons subjected to a space-time dependent velocity field. Next, we utilize these equations to compute the local density of current (LDC) following the general method proposed in Ref. \cite{dong2020berry}. The local density of a physical observable provides a convenient way to compute that observable. Given that orbital magnetization can induce current, the LDC can provide insights into the distribution of orbital magnetization, which is the primary focus of this work.

Let us consider electrons in periodic crystals governed by the Hamiltonian $H_0 \left(\hat{\mathbf{k}},\mathbf{r}\right)$, influenced by a space-time dependent velocity field $\mathbf{v}(\mathbf{r},t)$ that is significantly smaller than typical band velocities. Here, $\hat{\mathbf{k}}\equiv-i\boldsymbol{\nabla}_{\mathbf{r}}$ is the continuum momentum operator and has eigenvalue $\mathbf{k}$. 
As shown in Appendix \ref{app:A}, the effect of the velocity field can be captured by a perturbed Hamiltonian $H_0 \left(\hat{\mathbf{k}},\mathbf{r}\right)-\hat{\mathbf{k}}\cdot \mathbf{v} \left(\mathbf{r},t\right)$ that describes dynamics in a frame moving with the fluid. Physically, this form is simply a consequence of $\hat{\mathbf{k}}$ being the generator of continuum translations and can be intuited as a Galilean boost.

To derive the LDC, we introduce an auxiliary background momentum field $\mathbf{K}(\mathbf{r}, t)$, which shifts $\mathbf{k}\to\mathbf{k}-\mathbf{K}$ and is conjugate to the number current operator $\hat{\mathbf{J}} \equiv \nabla_{\hat{\mathbf{k}}} H_0$. In $U(1)$ gauge theories, $\mathbf{K}$ simply corresponds to the gauge field multiplied by the gauge charge. However, a background momentum is more general concept, essentially defining a spacetime-dependent reference point in momentum space about which the dynamics occurs, and is well-defined for neutral particles too. 

Mimicking gauge theories, this introduces an additional perturbation $-\hat{\mathbf{J}} \cdot \mathbf{K}(\mathbf{r}, t)$ into the Hamiltonian. We emphasize that the auxiliary momentum field $\mathbf{K}$ is introduced solely to calculate the LDC and determine the orbital magnetization. As is standard for response theory calculations, it will be taken to zero once the response functions are formally setup through appropriate functional derivatives. The total Hamiltonian is now 
%, where $\hat{\mathbf{V}}$ denotes the current operator
\begin{equation}
%\label{eq:ger9}
\hat{H}_{\text{tot}} =H_0 \left(\hat{\mathbf{k}},\mathbf{r}\right)-\hat{\mathbf{k}}\cdot \mathbf{v} \left(\mathbf{r},t\right)-\hat{\mathbf{J}}\cdot\mathbf{K}(\mathbf{r},t).
\end{equation}
Using the standard semiclassical wave-packet method (see Ref. \cite{sundaram1999wave} and Appendix \ref{app:B} for details), and assuming that both $\mathbf{v}$ and $\mathbf{K}$ vary slowly in space and time, the action for the wave packet takes the form $S = \int L dt$, with the Lagrangian $L$ given by 
\begin{equation}
L  =  \mathcal{A}_t+\dot{\mathbf{r}}_c\cdot\mathcal{A}_{\mathbf{r}_c}+\dot{\mathbf{k}}_c\cdot\mathcal{A}_{\mathbf{k}_c} +\dot{\mathbf{k}}_c \cdot \mathbf{r}_c- E.
\end{equation}
where $\mathbf{r}_c$ and $\mathbf{k}_c$ are the position and momentum of the center of mass of the wave packet. Additionally, $\mathcal{A}_{\alpha}=\langle u_n| i\partial_{\alpha} |u_n\rangle$ for $\alpha\in\{\mathbf{k}_c,\mathbf{r}_c,t\}$ represents the Berry connections, and the energy $E$ is

\begin{eqnarray}
%\label{eq:ger9.99}
  E &=& \epsilon_n+ \text{Im} \langle \partial_{k_i}u_n|\epsilon_n-\hat{H}|\partial_{\mathbf{v}}u_n\rangle\cdot\partial_{r_i}\mathbf{v} \nonumber\\
      &+&\text{Im} \langle \partial_{k_i}u_n|\epsilon_n-\hat{H}|\partial_{\mathbf{K}}u_n\rangle\cdot\partial_{r_i}\mathbf{K}.
  \end{eqnarray}
where the Hamiltonian $\hat{H} = H_0(\hat{\mathbf{k}}, \mathbf{r}) - \hat{\mathbf{k}} \cdot \mathbf{v}(\mathbf{r}_c, t) - \hat{\mathbf{J}} \cdot \mathbf{K}(\mathbf{r}_c, t)$ is obtained by evaluating the $\mathbf{v}$ and $\mathbf{K}$ in $\hat{H}_{\text{tot}}$ at the center of mass position $\mathbf{r}_c$ of the wave packet, and $\epsilon_n$ and $|u_n\rangle$ represent the $n$th eigenvalue and eigenstate of $\hat{H}$, respectively. Summation over the repeated index $i$ is implied henceforth and to simplify the notation, we will omit the band index $n$ and the index $c$ in $\mathbf{r}_c$. Since the wave packet size is much larger than the unit cell, we effectively enter the continuum limit. Consequently, we will replace $\mathbf{k}_c$ with $\mathbf{k}$ in the following discussion.

To obtain the EM, we apply the Euler-Lagrange equations to the Lagrangian $L$. The EM are given by:
\begin{equation}
%\label{eq:ger8.9}
\begin{cases}
\dot{\mathbf{r}}=\partial_{\mathbf{k}}E-\Omega_{\mathbf{k} t}-\Omega_{\mathbf{k} \mathbf{r}}\cdot \dot{\mathbf{r}}-\Omega_{\mathbf{k} \mathbf{k}}\cdot\dot{\mathbf{k}},\\ 
\dot{\mathbf{k}}= -\partial_{\mathbf{r}}E+\Omega_{\mathbf{r} t}+\Omega_{\mathbf{r} \mathbf{r}}\cdot\dot{\mathbf{r}}+ \Omega_{\mathbf{r}\mathbf{k}}\cdot\dot{\mathbf{k}}.
\end{cases}
\end{equation}
Here, $\Omega^{ij}_{\alpha\beta} = \partial_{\alpha_i} \mathcal{A}^j_{\beta} - \partial_{\beta_j} \mathcal{A}^i_{\alpha}$ represents the Berry curvature in the generalized space spanned by ${\mathbf{r}, \mathbf{k}, t}$.

The LDC, see Eq. (10) in the supplementary material of Ref. \cite{dong2020berry}, can be expressed as:
%\begin{eqnarray}
%\label{eq:led}
 %\rho^{\mathbf{V}}_{loc} &=& \lim_{\mathbf{K}\rightarrow0}\{-\int d\mathbf{k} D f [\frac{\partial E}{\partial\mathbf{K}}  -\Omega_{\mathbf{K}\mathcal{T}}]\nonumber\\
      %&-&\boldsymbol{\nabla}_{\mathbf{r}}\cdot\int d\mathbf{k} D f \text{Im} \langle \partial_{\mathbf{k}}u|\epsilon-\hat{H}|\partial_{\mathbf{K}}u\rangle\},
 % \end{eqnarray}  
  \begin{eqnarray}
\label{eq:led}
 \rho^{\mathbf{J}}_{\text{loc}} &=& \lim_{\mathbf{K}\rightarrow0} \left\{ -\int d\mathbf{k} \, D f \left[ \frac{\partial E}{\partial\mathbf{K}}  - \Omega_{\mathbf{K}\mathcal{T}} \right] \right. \nonumber\\
      &-& \left. \boldsymbol{\nabla}_{\mathbf{r}} \cdot \int d\mathbf{k} \, D f \, \text{Im} \left\langle \partial_{\mathbf{k}}u \left| \epsilon - \hat{H} \right| \partial_{\mathbf{K}}u \right\rangle \right\},
\end{eqnarray}
where $\int d\mathbf{k} \equiv \int \frac{d\mathbf{k}}{\left(2\pi\right)^\text{d}}$, with $\text{d}$ being the spatial dimension of the system, and $\Omega_{\mathbf{K}\mathcal{T}} = \dot{\mathbf{k}} \cdot \Omega_{\mathbf{Kk}} + \dot{\mathbf{r}} \cdot \Omega_{\mathbf{Kr}} + \Omega_{\mathbf{K}t}$, $D \equiv 1 + \Omega_{k_i r_i}$ is a modified phase space measure and the local equilibrium distribution is given by $f=\frac{1}{1+e^{\beta(\epsilon-\mathbf{k}\cdot\mathbf{v})}}$, where $\beta = T^{-1}$ is the inverse temperature. By performing integration by parts and using the EM, at constant temperature and chemical potential, the LDC reduces to (see Appendix \ref{app:B} for details)
\begin{eqnarray}
%\label{eq:ger9.99}
   \rho^{\mathbf{J}}_{\text{loc}} &=& 
      -\boldsymbol{\nabla}_{\mathbf{r}} \cdot \int d\mathbf{k} (-g\Omega_{ \mathbf{kk}}+f \text{Im} \langle \partial_{\mathbf{k}}u|\epsilon-\hat{H}|\partial_{\mathbf{k}}u\rangle) \nonumber\\
     &+& \int d \mathbf{k} f \Omega_{\mathbf{k}t} \equiv -\partial_i M^{i\mathbf{J}}  + \int d \mathbf{k} f \Omega_{\mathbf{k}t},
  \end{eqnarray}
 
Observe that for a time-dependent external field, the LDC includes contributions from the Berry curvature $\Omega_{\mathbf{k}t}$. This term accounts for the charge pumping current \cite{23}. For a filled band, using a gauge where $\mathcal{A}_t$ is periodic in the Brillouin zone, the last term in $\rho^{\mathbf{J}}_{\text{loc}}$ is formulated as:
\begin{equation}
\int d \mathbf{k} \Omega_{\mathbf{k}t} = -\partial_t \int d \mathbf{k} \mathcal{A}_{\mathbf{k}} = -\partial_t \mathbf{P},
\end{equation}
where $\mathbf{P}$ is the polarization. Thus, this term corresponds to the current induced by polarization.
%On the other hand, t 
The quantity 
\begin{equation}
%\label{eq:ger9.9}
M^{iJ_j} = \int d\mathbf{k} (-g\Omega_{k_ik_j}+f \text{Im} \langle \partial_{k_i}u|\epsilon-\hat{H}|\partial_{k_j}u\rangle) \equiv M_{ij},
\end{equation}
represents the dipole density tensor of current or the orbital magnetization tensor \cite{dong2020berry}. Its vector form is
\begin{equation}
M^{\text{orb}}_{l}=\frac{\epsilon_{lij}M_{ij} }{2}= \int d\mathbf{k} (f m^{\text{orb}}_l - g\Omega_{l}\left(\mathbf{k}\right)), 
\end{equation}
where $\Omega_{l}(\mathbf{k})$ is the $l$-th component of the Berry curvature, and $m_{l}^{\text{orb}}$ is the $l$-th component of the orbital magnetic moment. The function $g$ is given by:
\begin{equation}
g=-\frac{1}{\beta}\ln[1+e^{-\beta(\epsilon-\mathbf{k}\cdot\mathbf{v})}].
\end{equation}

Finally, by reintroducing the band index and summing the contributions of each band, we obtain the density of orbital magnetization
 \begin{equation}
\mathbf{M}^{\text{orb}}= \sum_n\int d\mathbf{k} \{f \mathbf{m}_n^{\text{orb}} +\frac{1}{\beta}\ln[1+e^{-\beta(\epsilon_{n,\mathbf{k}}-\mathbf{k}\cdot\mathbf{v})}]\boldsymbol{\Omega}_n\}.
\end{equation}
%The above equation is one of the main results of this work. 
\begin{table*}[!]
    \centering
    \begin{tabular}{c|c|c|c|c}
    \hline
    Limit & Definition &$\chi_{ij}^{\text{occ}}$ & $\chi_{ij}^{\text{Fs}}$& $\chi^\text{Weyl}_{ij}=\chi^{\text{occ}}_{ij}+\chi^{\text{Fs}}_{ij}$\\[6pt]
    \hline
    \hline
    Uniform, clean & $ \tilde{v} q \ll |\omega|, 1\ll |\omega|\tau, \text{arbitrary}\ \tilde{v} q\tau$ & $\sum_{n}\int_{k}\Omega_{n,i}(\mathbf{k})\Theta(-\epsilon_{n,\mathbf{k}})k_{j}$ & $ 0$&$\frac{1}{6}\chi^{\mathcal{C}}_{ij}=\frac{1}{6}\chi^{\mathcal{C}}_{ij}+0$\\[6pt]
    \hline
    Static, clean & $ |\omega| \ll \tilde{v} q, 1\ll \tau \tilde{v} q, \text{arbitrary}\ \omega\tau$ &$\sum_{n}\int_{k}\Omega_{n,i}(\mathbf{k})\Theta(-\epsilon_{n,\mathbf{k}})k_{j}$ & $ {\sum_{n}\int_{{k}} m_{n,i}^\text{orb}(\mathbf{k})\delta(\epsilon_{n,\mathbf{k}})k_{j}}$& $\frac{1}{2}\chi^{\mathcal{C}}_{ij}=\frac{1}{3}\chi^{\mathcal{C}}_{ij}+\frac{1}{6}\chi^{\mathcal{C}}_{ij}$ \\[6pt]
    \hline
    Uniform, disorder & $\tilde{v} q\tau\ll |\omega|\tau\ll1 $ &$\sum_{n}\int_{k}\Omega_{n,i}(\mathbf{k})\Theta(-\epsilon_{n,\mathbf{k}})k_{j}$ & $ {\sum_{n}\int_{{k}} m_{n,i}^\text{orb}(\mathbf{k})\delta(\epsilon_{n,\mathbf{k}})k_{j}}$& $\frac{1}{2}\chi^{\mathcal{C}}_{ij}=\frac{1}{3}\chi^{\mathcal{C}}_{ij}+\frac{1}{6}\chi^{\mathcal{C}}_{ij}$ \\[6pt]
    \hline
    Static, disorder & $ |\omega|\tau \ll \tilde{v} q\tau \ll1$ & $\sum_{n}\int_{k}\Omega_{n,i}(\mathbf{k})\Theta(-\epsilon_{n,\mathbf{k}})k_{j}$ & $ {\sum_{n}\int_{{k}} m_{n,i}^\text{orb}(\mathbf{k})\delta(\epsilon_{n,\mathbf{k}})k_{j}}$&$\frac{1}{2}\chi^{\mathcal{C}}_{ij}=\frac{1}{3}\chi^{\mathcal{C}}_{ij}+\frac{1}{6}\chi^{\mathcal{C}}_{ij}$\\[6pt]
    \hline
    \end{tabular}
    \caption{Summary of results as $q\to0$ and $\omega\to0$ at various orders is presented for general band structures. In the definition column, $\tilde{v}\equiv|\nabla_{\mathbf{k}}\epsilon_{n,\mathbf{k}}|$. The last column represents the orbital magnetization for an isotropic Weyl fermion with a velocity of $\tilde{v}=v_F$, chiral charge $\mathcal{C}$, and chemical potential $\mu$ relative to the Weyl node. Here, $\chi^{\mathcal{C}}_{ij}=\mathcal{C}\left(\frac{\mu}{2\pi v_F}\right)^2\delta_{ij}$.}
    \label{tab:Results}
\end{table*}

For a parabolic band with Berry phases due to internal degrees of freedom, the same equation for orbital magnetization can be obtained using the free energy density (see Appendix \ref{app:C} for details).
The converse vortical effect refers to the response of orbital magnetization to $\mathbf{v}$ in linear order. Formally Taylor expanding in $\mathbf{v}$ leads to:
%%{\cyan{What do the internal degrees of freedom refer to? The Berry phase can exist for an isolated band.}} 
\begin{align}
\mathbf{M}^\text{orb}(\mathbf{v}) 
&= \sum_{n}{\int_{\mathbf{k}}\mathbf{m}^\text{orb}_{n} f } \nonumber \\
&\quad + \frac{1}{\beta}\int_{\mathbf{k}}\boldsymbol{\Omega}_{n}\ln\left(1+e^{-\beta\left(\epsilon_{n,\mathbf{k}}-\mathbf{k\cdot v}\right)}\right) \nonumber \\
&\equiv \mathbf{M}_0^\text{orb}+\chi^\text{orb}{\cdot \mathbf{v}}+O\left(\mathbf{v^{2}}\right),
\end{align}
where the $\mathbf{v}$-independent term $\mathbf{M}_0^\text{orb}$ denotes the intrinsic equilibrium magnetization and the tensor $\chi^\text{orb}$ is the desired response function for the converse vortical effect. At $T=0$, $f=\Theta(-\epsilon_{n,\mathbf{k}}+\mathbf{k\cdot v})$, and $\chi^\text{orb}$ reduces to:
\begin{align}
\label{eq:chi_stat}
\chi_{ij}^\text{orb} &= {\sum_{n}\int_{\mathbf{k}} m_{n,i}^\text{orb}\delta(\epsilon_{n,\mathbf{k}})k_{j}} \nonumber \\
&\quad + \sum_{n}\int_{\mathbf{k}}\Theta(-\epsilon_{n,\mathbf{k}})\Omega_{n,i}k_{j}\equiv \chi_{ij}^\text{Fs} +\chi_{ij}^\text{occ}.
\end{align}

The equation above reveals that the magnetic susceptibility is determined by the orbital magnetic moment of electrons on the Fermi surface (indicated as $\chi_{ij}^\text{Fs} $) as well as the Berry curvature of the occupied bands (indicated as $\chi_{ij}^\text{occ}$). Interestingly, this response function takes the same form as the vortical effect \cite{nanda2023vortical}. However, they represent distinct responses. The vortical effect is defined as the axial current response to angular velocity while the converse vortical effect is the orbital magnetization response to linear velocity.}}

\section{Kubo Formula for Converse Vortical Effect}\label{sec:Kubo}

The Kubo formula provides a fundamental theoretical framework for analyzing the linear response of a system to external perturbations. In this section, we apply the Kubo formula to compute $\chi_{ij}^\text{orb}$ for a space-time-dependent velocity field, considering general values of $\mathbf{q}$ and $\omega$, and incorporating the effects of quasiparticle lifetime $\tau$. The motivation for using the Kubo formula lies in its ability to elucidate the system's response across different limits (e.g., static, uniform), providing valuable insights into the fundamental physical mechanisms governing the orbital magnetization. 
 
%Similar to the vortical effects \cite{nanda2023vortical}, we analyze the static limit (cCVE) and uniform limit (cGVE) separately to gain a comprehensive understanding of the behavior of orbital magnetization in different regimes. 
The Bloch energy and wave function for the $n$-th band are given by $\epsilon_{n,\mathbf{k}}$ and $\psi_{n,\mathbf{k}}(\mathbf{r})=\psi_{n,\mathbf{k}}(\mathbf{R}+\boldsymbol{\rho})=N^{-1/2}e^{i\mathbf{k}\cdot(\mathbf{R}+\boldsymbol{\rho})}u_{n,\mathbf{k}}(\boldsymbol{\rho})$, respectively, where $\mathbf{k}$ represents the Bloch momentum of electrons, and $\mathbf{R}$ denotes the coordinates of the unit cells, $\boldsymbol{\rho}$ represents position within each unit cell and $N$ is the total number of unit cells. 
In this basis, the matrix elements of the velocity-induced perturbation $H_1 =-\hat{\mathbf{k}}\cdot\mathbf{v}\left(\mathbf{r},t\right)$ with $\hat{\mathbf{k}}$ denoting the momentum operator, are $\langle \psi_{n,\mathbf{k}}|H_1|\psi_{m,\mathbf{k+q}}\rangle = (2\pi)^3\langle u_{n,\mathbf{k}}|(\mathbf{k}-i \nabla_{\boldsymbol{\rho}})|u_{m,\mathbf{k+q}}\rangle\cdot \mathbf{v}(\mathbf{q},t)$, where $\mathbf{v}(\mathbf{q},t)$ is $\mathbf{v}(\mathbf{r},t)$ Fourier transformed to momentum space. 
Thus, it is convenient to introduce the operator $\hat{\mathbf{Q}}=\mathbf{k}-i \nabla_{\boldsymbol{\rho}}$ and write $H_1= -\hat{\mathbf{Q}}\cdot\mathbf{v}$. Unlike the continuum perturbation $-\mathbf{k}\cdot\mathbf{v}$, $H_1$ respects the Brillouin Zone periodicity and can be viewed as the kinematic analog of minimal coupling $\mathbf{J}\cdot\mathbf{A}$ that is well-defined on a lattice through Peierl's substitution \cite{nanda2023vortical}.

To calculate $\chi_{ij}^\text{orb}$, we Fourier transform $\mathbf{M}^\text{orb}=\frac{1}{2}\mathbf{r}\times \mathbf{J}$ to Bloch momentum and Matsubara frequencies, $\mathbf{M}^\text{orb}(\mathbf{q},iq_{n})=\frac{i}{2}\nabla_{\mathbf{q}}\times \mathbf{J}(\mathbf{q},iq_{n})$, and compute the susceptibility, $\chi_{ij}^\text{orb}(\mathbf{q},iq_{n})=\frac{\partial M^\text{orb}_{i}}{\partial v_{j}}(\mathbf{q},iq_{n})$ with $i$ and $j$ denoting spatial components. The basic one-loop diagram yields
%Operationally, we use the Kubo formula to calculate the linear susceptibility for a current density $J(\mathbf{q},iq_{n})$ to linear order in $\mathbf{v}(\mathbf{q),iq_n)$. Substituting this current density back into the above definition for the orbital magnetization, we derive the following expression for the magnetic susceptibility_
\begin{align}
\chi_{ij}^\text{orb}(\mathbf{q},iq_{n}) &= -\epsilon_{i\mu\nu}i\partial_{q_{\mu}}\frac{1}{2\beta}\sum_{i\nu_{n}}\intop_{\mathbf{k}}\nonumber \\
&\text{tr}\left[J_{\nu}(\mathbf{k}+\mathbf{q})G_{0}(\mathbf{k},i\nu_{n})G_{0}(\mathbf{k}+\mathbf{q},i\nu_{n}+iq_{n})Q_j\right]
\end{align}
where $G_{0}(\mathbf{k}, i\nu_{n}) = \left[i\nu_{n} - H_{0}(\mathbf{k})+i\text{sgn}(\nu_n)/2\tau\right]^{-1}$
is the unperturbed Matsubara Green's function, the elements of the  matrix $Q_j$ are $Q^{mn}_j=\langle u_{m,\mathbf{k}} |\hat{Q}_j| u_{n,\mathbf{k}+\mathbf{q}}\rangle$, $J_{\nu}(\mathbf{k}) = \frac{\partial H_{0}(\mathbf{k})}{\partial k_{\nu}}$ is the current operator and repeated indices are summed. The retarded response function follows from analytically continuing $iq_{n} \rightarrow \omega + i0^{+}$. %Equation (8) describes the response of the orbital magnetization of Bloch electrons in a general lattice to an external velocity field at momentum $\mathbf{q}$ and frequency $\omega$. 
The Matsubara sum yields
\begin{align}
\chi_{ij}^\text{orb}(\mathbf{q},iq_{n}) &= -\frac{1}{2}\epsilon_{i\mu\nu}i\partial_{q_{\mu}}\int_{\boldsymbol{k}}\sum_{n,m}S_{m,n}(\mathbf{k},\mathbf{q},iq_{n}) \nonumber \\
&\left\langle u_{n,\mathbf{k}+\mathbf{q}}\left|J_{\nu}(\mathbf{k}+\mathbf{q})\right|u_{m,\mathbf{k}}\right\rangle Q^{mn}_j,
\end{align}
%(\mathbf{k},\mathbf{q},iq_{n})
where
\begin{align}
 S_{m,n}(\mathbf{k},\mathbf{q},iq_{n}) = \frac{1}{\beta}\sum_{i\nu_{n}}\frac{1}{i\nu_{n}-\epsilon_{m,\mathbf{k}}+i\frac{\text{sgn}(\nu_{n})}{2\tau}} \nonumber \\
 \frac{1}{i\nu_{n}+iq_{n}-\epsilon_{n,\mathbf{k}+\mathbf{q}}+i\frac{\text{sgn}(\nu_{n}+q_{n})}{2\tau}}.
\end{align}
%, as detailed in \cite{nanda2023vortical}
At zero temperature, the off-diagonal elements of the matrix $Q_j$ couple to the inter-band orbital magnetization matrix of the Bloch electrons \cite{ogata2017theory}, contributing to the orbital magnetization (for more details, see the Appendix \ref{app:D}). 

We now focus on the zero-temperature regime in the context of both the nearly-free electron approximation and the deep tight-binding approximation \cite{nanda2023vortical}. In the deep tight-binding approximation, within the unit cell, the lattice potential can be approximated by the form $V\left(\boldsymbol{\rho}\right) = \sum_j V_j \delta\left(\boldsymbol{\rho} - \boldsymbol{\rho}_j\right)$, where $\boldsymbol{\rho}_j$ and $\boldsymbol{\rho}$ denote discrete points and the continuous position variable within a unit cell, respectively. The Bloch function $u_{n,\mathbf{k}}\left(\boldsymbol{\rho}\right) = \sum_j u_{n,\mathbf{k}}\phi_{j}\left(\boldsymbol{\rho}\right)$, where $\mathbf{k}$ is within the first Brillouin zone, and the function $\phi_{j}\left(\boldsymbol{\rho}\right) \approx e^{-\sqrt{2|E|m} |\boldsymbol{\rho}-\boldsymbol{\rho}_j|}$, with $E$ and $m$ denoting the energy and mass of the electron, respectively. The term $\langle\phi_i |i\nabla{\boldsymbol{\rho}}|\phi_j\rangle\ll 1$ if $i\neq j$ and $\langle\phi_j|i\nabla_{\boldsymbol{\rho}}|\phi_j\rangle=0$ since $\phi_j\left(\boldsymbol{\rho}\right)$ have definite parity, hence, the inner product $\langle u_{n,\mathbf{k}}|i\nabla_{\boldsymbol{\rho}}|u_{n,\mathbf{k^{'}}}\rangle$ is exponentially small. Under these conditions, the Fermi distribution function $f(\epsilon_{n,\mathbf{k}})$ and $Q^{mn}_j$ simplify to $\Theta(-\epsilon_{n,\mathbf{k}})$ and $k_j\delta_{mn}$, respectively. The difference between various orders of limits of $\omega\to0$, $q\to0$ and $\tau\to\infty$ is determined by the behavior of $S_{m,n}$ in these limits. In the static limit ($\omega\to0$ followed by $q \rightarrow 0$), we find $\chi_{ij}^\text{orb}$ reduces to Eq. (\ref{eq:chi_stat}) derived using semiclassical chiral kinematic theory for both $v_F q\tau\gg1$ and $v_F q\tau\ll1$, where $v_F$ is a typical band velocity. In contrast, the dirty uniform limit, ($\mathbf{q} \rightarrow 0$ followed by $\omega \rightarrow 0$ with $|\omega\tau|\ll1$), leads to Eq. (\ref{eq:chi_stat}) while the clean uniform limit ($\mathbf{q} \rightarrow 0$ followed by $\omega \rightarrow 0$ with $|\omega\tau|\gg1$) gives:
\begin{equation}
\label{eq:chi_uniform}
\chi_{ij}^\text{orb}=\intop_{\mathbf{k}}\sum_{n}\Theta\left(-\epsilon_{n,\mathbf{k}}\right)\Omega_{n,i}(\mathbf{k})k_{j}
\end{equation}
Thus, $\chi_{ij}^\text{orb}$ in this limit is solely determined by $\boldsymbol{\Omega}_n(\mathbf{k})$ of occupied bands. It is worth noting that the $\boldsymbol{\Omega}_n(\mathbf{k})$ contribution vanishes for a filled band in the continuum limit at zero temperature, where $\hat{Q}_j\to \hat{k}_j$ and $u_{n,\mathbf{k}}$ become $\mathbf{k}$-independent as $k\to\infty$, similarly to the GVE \cite{nanda2023vortical}. Therefore, only partially filled bands contribute to $\mathbf{M}^\text{orb}$ in any limit. The results are summarized in Table \ref{tab:Results}. However, in a disordered electron fluid with relatively short $\tau$, both the Berry curvature of the occupied bands and the orbital moment of electrons on the Fermi surface contribute to the magnetic susceptibility in both the static and uniform limits. This magnetic susceptibility takes the same form as described in Eq. (\ref{eq:chi_stat}), acquiring the combined effects of the Berry curvature and the orbital moment. In analogy with the CVE and the GVE, we refer to the two results for the converse effects, Eqs. (\ref{eq:chi_stat}) and (\ref{eq:chi_uniform}) as the cCVE and the cGVE, respectively.

\section{cCVE and cGVE of Weyl fermions}\label{sec:Weyl}
We now evaluate $\chi_{ij}^\text{orb}$ for a single, isotropic, continuum Weyl fermion with chirality $\mathcal{C}=\pm1$. The effective Hamiltonian is given by $H(\mathbf{k})=\mathcal{C}\mathbf{k\cdot\boldsymbol{\sigma}-\mu-k\cdot v}$, where $\boldsymbol{\sigma}$ represents the Pauli matrices and $\mu$ is the chemical potential relative to the Weyl node and $\mathbf{v}$. The results at $T=0$ are stated in Table \ref{tab:Results}. Since the effect is proportional to the chirality $\mathcal{C}$ and $\mu^2$, which are odd and even under improper symmetries, respectively, all improper symmetries must be broken for a material with pairs of Weyl fermions, rendering the material chiral overall in addition to each Weyl node being chiral.

\section{Experimental realization}\label{expt}
To experimentally observe the converse vortical effects, we propose a simple experiment sketched in Fig. \ref{Fig.1}. This is significantly simpler than the curved geometries required for the vortical effects  \cite{nanda2023vortical}. By leveraging a temperature difference gradient ($\nabla T \approx 1K/\mu m$) and a Seebeck coefficient ($S=100a \mu V/K$), we generate an electric field strength of $|\mathbf{E}|=0.1V/m$, driving the motion of electrons relative to the lattice. Consequently, $\mathbf{M}^\text{orb}$ aligns with $\mathbf{v}=\mu_{\mathrm{mob}}\mathbf{E}$, with $\mu_{\mathrm{mob}}$ representing the mobility of the system. For a WSM with typical parameter values such as $\mu_{\mathrm{mob}}=10^{5}cm^{2}/(Vs)$, $v_{F}=10^{5}m/s$, and Fermi energy differences $\mu_{\pm}=\left(0.5\pm0.025\right)eV$ relative to the left-handed/right-handed Weyl nodes, $|\mathbf{M}^\text{orb}|\approx4.68\times10^{-2}A/m$. Moreover, Weyl nodes are not mandatory, and the converse vortical effects can also occur in a chiral semimetal such as CoSi \cite{rao2019observation,takane2019observation,xu2020optical}.

\section{Summary}

We employed the chiral kinematic theory for investigating the influence of space-time dependent velocity fields on electron fluids. Through analysis of the LDC, we explore the orbital magnetization response, that we dub the converse vortical effect, induced by the velocity field. By applying the Kubo formula, we calculate the converse vortical effect under different limits. Our study reveals that the magnetic susceptibility in the static limit, which encompasses both clean and disordered systems, and in the uniform limit of disordered systems, is determined by the orbital moment on the Fermi surface and the Berry curvature of occupied bands. These findings are in agreement with the predictions derived from chiral kinematic theories. However, in the uniform limit of clean systems, the susceptibility is solely determined by the Berry curvature of occupied bands. This research provides valuable insights into the behavior of electron fluids under space-time-dependent velocity fields, shedding light on the intricate relationship between the velocity field and electron properties. Our results contribute to advancing the understanding of fundamental physical phenomena and offer opportunities for exploring new applications in electron fluid systems.

\acknowledgments

We acknowledge support from the Department of Energy grant no. DE-SC0022264.
\bibliographystyle{unsrt}
\bibliography{ivelib}

\appendix
\onecolumngrid

%\section{Supplemental Material for Chiral kinematic theory and converse vortical effects}
\makeatletter
\renewcommand\section{\@startsection{section}{1}{\z@}%
  {-3.5ex \@plus -1ex \@minus -.2ex}%
  {2.3ex \@plus.2ex}%
  {\normalfont\LARGE\bfseries}}

\renewcommand\subsection{\@startsection{subsection}{2}{\z@}%
  {-3.05ex\@plus -1ex \@minus -.2ex}%
  {1.05ex \@plus .2ex}%
  {\normalfont\Large\bfseries}}
\makeatother

\section{Perturbation induced by velocity field}   
\label{app:A}

In this section, we derive the perturbation term $-\hat{\mathbf{k}} \cdot \mathbf{v}\left(\mathbf{r}, t\right)$ induced by a space-time dependent velocity field $\mathbf{v}\left(\mathbf{r}, t\right)$. 
To achieve this, we consider electrons on a lattice governed by a Bloch Hamiltonian $H_{0}(\hat{\mathbf{k}})$. Under the influence of a space-time dependent velocity field, the position of the electrons is shifted by a small space-time dependent displacement $\mathbf{x}(\mathbf{r}, t)$ relative to the lattice background. {\red{Thus, in a reference frame moving with the electron fluid, the the wave function evolves in time as}}

\begin{equation}
\label{eq:boost0}
\psi\left(\mathbf{r},t\right)=\hat{T}[e^{i\int_{\mathbf{x}\left(0\right)} ^{\mathbf{x}\left(t\right)}  \hat{\mathbf{k}}\cdot d \mathbf{x}\left(\mathbf{r},\tau\right)-i t H_{0}\left(\hat{\mathbf{k}}\right)}]\psi\left(\mathbf{r},0\right)=\hat{T}[e^{i\int^{t}_{0}d\tau \hat{\mathbf{k}}\cdot \frac{\partial \mathbf{x}\left(\mathbf{r},\tau\right)}{\partial \tau}-i t H_{0}\left(\hat{\mathbf{k}}\right)}]\psi\left(\mathbf{r},0\right), 
\end{equation}
where $\hat{T}$ denotes time ordering and $\hat{\mathbf{k}}$ is the momentum operator that generates spatial translations. By factoring out the "unperturbed time-dependence", we can derive the perturbation term by defining $\psi_I(\mathbf{r}, t) = e^{i t H_{0}(\hat{\mathbf{k}})} \psi(\mathbf{r}, t)$, which leads to:
\begin{equation}
i\partial_t \psi_I\left(\mathbf{r},t\right) = -\hat{\mathbf{k}}\cdot\partial_t\mathbf{x}\left(\mathbf{r},t \right)\psi_I\left(\mathbf{r},t\right)\equiv-\hat{\mathbf{k}}\cdot\mathbf{v}\left(\mathbf{r},t \right)\psi_I\left(\mathbf{r},t\right).
\end{equation}
The above equation indicates that $\psi_I\left(\mathbf{r},t\right)$ behaves like a wave function in the Interaction Picture with an unperturbed Hamiltonian $H_{0}\left(\hat{\mathbf{k}}\right)$ and perturbation $-\hat{\mathbf{k}}\cdot\mathbf{v}\left(\mathbf{r}, t\right)$. Therefore, the total Hamiltonian {\red{in the moving frame}} can be expressed as:
\begin{equation}
\label{eq:boost1}
H\left(\hat{\mathbf{k}},\mathbf{r},t \right) = H_{0}\left(\hat{\mathbf{k}}\right)-\hat{\mathbf{k}}\cdot\mathbf{v}\left(\mathbf{r}, t\right).
\end{equation}

%It is worth noting that the velocity field is an external field given
%by $\mathbf{v}(\mathbf{r},t)=\frac{d\mathbf{x}(\mathbf{r},t)}{dt}$,
%where $\mathbf{x}(\mathbf{r},t)$ is the displacement of electrons
%induced by external sources such as temperature gradients or rotation.
%A concrete example is rotation, which
%induces an additional space-time dependence into the time evolution
%of any wave function: $\psi\left(\mathbf{r},t\right)=\hat{T}[e^{i\int_{0}^{t}d\tau\hat{\mathbf{L}}\cdot\boldsymbol{\Omega}\left(\mathbf{r},\tau\right)-itH_{0}\left(\hat{\mathbf{k}}\right)}]\psi\left(\mathbf{r},0\right)$,
%where $\hat{\mathbf{L}}=\mathbf{r}\times\hat{\mathbf{k}}$ and $\boldsymbol{\Omega}\left(\mathbf{r},\tau\right)$
%are the angular momentum operator and the angular velocity, respectively. The term $\hat{\mathbf{L}}\cdot\boldsymbol{\Omega}\left(\mathbf{r},\tau\right)$
%can be expressed as $\hat{\mathbf{L}}\cdot\boldsymbol{\Omega}\left(\mathbf{r},\tau\right)=\hat{\mathbf{k}}\cdot\mathbf{v}\left(\mathbf{r},\tau\right)$ with the velocity field $\mathbf{v}\left(\mathbf{r},\tau\right)=\boldsymbol{\Omega}\left(\mathbf{r},\tau\right)\times\mathbf{r}$.

{\blue{The perturbation term $-\hat{\mathbf{k}} \cdot \mathbf{v}$ is naturally associated with the Galilean boost. For instance, for electrons described by the Hamiltonian $H_0=\frac{\hat{\mathbf{k}}^2}{2m}$, the momentum $\hat{\mathbf{k}}$ transforms as $\hat{\mathbf{k}} \rightarrow \hat{\mathbf{k}} - m\mathbf{v}$, and the Hamiltonian $H_0=\frac{\hat{\mathbf{k}}^2}{2m}$ is modified to $H=\frac{\left(\hat{\mathbf{k}} - m\mathbf{v}\right)^2}{2m} = \frac{\hat{\mathbf{k}}^2}{2m} - \hat{\mathbf{k}} \cdot \mathbf{v} + O\left(\mathbf{v}^2\right)$ \cite{anandan2004quantum}. However, the derivation of the perturbation term $-\hat{\mathbf{k}} \cdot \mathbf{v}$ in Eq. (\ref{eq:boost1}) is more general and not restricted to this specific context. For instance, for massless Weyl fermions, the transformation $\hat{\mathbf{k}}\to\hat{\mathbf{k}}-m\mathbf{v}$ fails because $m=0$, but our results in Eq. (\ref{eq:boost1}) remain valid. Indeed, the perturbation $-\mathbf{k}\cdot{\mathbf{v}}$ is routinely employed in the literature to describe Weyl fermions moving with constant linear velocity \citep{chen2014lorentz}. Eq. (\ref{eq:boost1}) merely generalizes it to smooth space-time dependent velocities.}}

\section{ Equations of semiclassical motion } 

\label{app:B}

%{\red{Split this section into two-three sub-sections, one for the recap and one-two for our work.}}
{\blue{In this section, we explore the semiclassical equation of motion using the semiclassical wave-packet dynamics method. First, we review the equation of motion for electrons under a general space-time dependent perturbation. Detailed discussions can be found in 
Ref. \cite{sundaram1999wave}. Secondly, we apply these equations of motion (EM) to a system under a space-time dependent velocity field and and derive the density of orbital magnetization.

%{\red{While recapping someone else's work, I suggest using third person "one does this, one does that, Ref 73 did so and so, etc." Avoid "we do this, we do that"}}
\subsection{ Semiclassical wave-packet dynamics } 
Consider a system under a set of general space-time dependent perturbations with the following Hamiltonian:
\begin{equation}
\label{eq:ger1}
H \equiv H\left(\hat{\mathbf{k}},\mathbf{r};\theta(\mathbf{r},t),t\right),
\end{equation}
where the first $\mathbf{r}$ represents the fast-changing part, and the $\theta(\mathbf{r},t)$ represents the set of slowly changing perturbations. After expanding the slowly changing part of the Hamiltonian around wave packet center $\mathbf{r}_c$, Ref. \cite{sundaram1999wave} obtained
\begin{equation}
\label{eq:ger2}
H = H_c\left(\hat{\mathbf{k}},\mathbf{r};\theta(\mathbf{r}_c,t),t\right) + H_1\left(\hat{\mathbf{k}},\mathbf{r};\mathbf{r}_c,t\right)+....
\end{equation}
%with first order term $H_1\equiv H_1\left(\hat{\mathbf{k}},\mathbf{r};\mathbf{r},\mathbf{r}_c,t\right)=\frac{1}{2}\partial_{\mathbf{r}_c}H_c \cdot\left(\mathbf{r}-\mathbf{r}_c\right)+h.c.$.

In general, the local Hamiltonian $H_c$ has eigenvalue $\epsilon_n\left( \mathbf{p},\mathbf{r}_c,t\right)$ and eigenstate $e^{i\mathbf{p}\cdot\mathbf{r}}|u_{n}\left(  \mathbf{p},\mathbf{r}_c, t \right)\rangle$. To obtain the equations of motion, Ref. \cite{sundaram1999wave} first derive the Lagrangian $L\equiv\langle \Psi| \left(i\partial_t -H\right)|\Psi\rangle$, where the wave packet wave function is constructed as a superposition of Bloch states from band $n$
\begin{equation}
\label{eq:ger3}
|\Psi\rangle=\int d\mathbf{p} a(\mathbf{p}) e^{i\mathbf{p}\cdot\mathbf{r}}|u_{n}\left(  \mathbf{p},\mathbf{r}_c, t \right)\rangle.
\end{equation}
The normalization of  $|\Psi\rangle$ indicates $\int d\mathbf{p} |a(\mathbf{p})| =1$, and the magnitude of $a(\mathbf{p})=|a(\mathbf{p})|e^{-i\gamma\left(\mathbf{p},\mathbf{r},t\right)}$ satisfies $|a(\mathbf{p})|^2\approx\delta(\mathbf{p}-\mathbf{k}_c)$, where $\mathbf{k}_c = \langle \Psi | \hat{\mathbf{k}}|\Psi\rangle$ is the center of mass of momentum of the wave packet. The center of mass position $\mathbf{r}_c$ of the wave packet is determined in the following way
\begin{equation}
\label{eq:ger3.0}
\mathbf{r}_c = \langle\Psi| \mathbf{r}|\Psi\rangle=\int d\mathbf{p}d\mathbf{q} a^*(\mathbf{p})a(\mathbf{q}) \langle u_{n}\left(  \mathbf{p},\mathbf{r}_c, t \right)|(i\partial_{\mathbf{p}}e^{-i\mathbf{p}\cdot\mathbf{r}})e^{i\mathbf{q}\cdot\mathbf{r}}|u_{n}\left(  \mathbf{q},\mathbf{r}_c, t \right)\rangle = \partial_{\mathbf{k}_c}\gamma\left(\mathbf{k}_c,\mathbf{r}_c,t\right) + \mathcal{A}_n \left(\mathbf{k}_c,\mathbf{r}_c,t\right) ,
\end{equation}
where $\gamma\left(\mathbf{k}_c,\mathbf{r}_c,t\right) = -\arg a(\mathbf{k}_c)$, and $\mathcal{A}_n \left(\mathbf{k}_c,\mathbf{r}_c,t\right)$ is the Berry connection in band $n$.
The energy part can be evaluated:
\begin{equation}
\label{eq:ger4.0}
\langle\Psi | H_c | \Psi\rangle = \epsilon_n\left( \mathbf{k}_c,\mathbf{r}_c,t\right),
\end{equation}
\begin{equation}
\label{eq:ger4}
%\langle\Psi | H_1 | \Psi\rangle = \sum_{m\neq n} Re[\langle u_n |\partial_{\mathbf{r}_c} H_c | u_m\rangle \langle u_m | i\partial_{\mathbf{k}_c} u_n\rangle],
\langle\Psi | H_1 | \Psi\rangle =  \text{Im} \langle \partial_{p^i_{c}}u_n|\epsilon-H_c|\partial_{\mathbf{\theta}}u_n\rangle\cdot\partial_{r^i_{c}}\mathbf{\theta},
\end{equation}
Additionally, the dynamic part takes the following form
\begin{equation}
\label{eq:ger5}
\langle \Psi| i\partial_t|\Psi\rangle = \langle u_n| i\partial_t |u_n\rangle+\dot{\mathbf{r}}_c\cdot\langle u_n| i\partial_{\mathbf{r}_c} |u_n\rangle+\dot{\mathbf{k}}_c\cdot\langle u_n| i\partial_{\mathbf{k}_c} |u_n\rangle - \mathbf{r}_c \cdot \dot{\mathbf{k}}_c,
\end{equation}
up to a global time derivative.
Therefore, the Lagrangian takes the from
\begin{equation}
\label{eq:ger6}
L  =  \langle u_n| i\partial_t |u_n\rangle+\dot{\mathbf{r}}_c\cdot\langle u_n| i\partial_{\mathbf{r}_c} |u_n\rangle+\dot{\mathbf{k}}_c\cdot\langle u_n| i\partial_{\mathbf{k}_c} |u_n\rangle - \mathbf{r}_c \cdot \dot{\mathbf{k}}_c- E_n\left(\mathbf{k}_c,\mathbf{r}_c,t\right),
\end{equation}
where the energy 
\begin{equation}
\label{eq:ger7}
E_n\left(\mathbf{k}_c,\mathbf{r}_c,t\right) = \epsilon_n\left( \mathbf{k}_c,\mathbf{r}_c,t\right)+ \text{Im} \langle \partial_{p_i}u_n|\epsilon-\hat{H}|\partial_{\mathbf{\theta}}u_n\rangle\cdot\partial_{r_i}\mathbf{\theta}\equiv E_n.
\end{equation}
Using the Lagrangian, one can derive the EM using the Euler-Lagrange equations. The EM is given by
\begin{equation}
\label{eq:ger8}
\begin{cases}
\dot{\mathbf{r}}_c=\partial_{\mathbf{k}_c}E_{n}-\Omega_{\mathbf{k}_c t}-\Omega_{\mathbf{k}_c \mathbf{r}_c}\cdot \dot{\mathbf{r}}_c-\Omega_{\mathbf{k}_c \mathbf{k}_c}\cdot\dot{\mathbf{k}}_c,\\ 
\dot{\mathbf{k}}_c= -\partial_{\mathbf{r}_c}E_{n}+\Omega_{\mathbf{r}_c t}+\Omega_{\mathbf{r}_c \mathbf{r}_c}\cdot\dot{\mathbf{r}}_c+ \Omega_{\mathbf{r}_c\mathbf{k}_c}\cdot\dot{\mathbf{k}}_c.
\end{cases}
\end{equation}
where $\Omega^{ij}_{\alpha\beta} = \partial_{\alpha_i} \mathcal{A}^j_{\beta} -\partial_{\beta_j} \mathcal{A}^i_{\alpha}$ is the Berry curvature in generalized space spanned by $\{\mathbf{r}_c,\mathbf{k}_c,t\}$, and $\mathcal{A}^j_{\alpha}\equiv \langle u_n| i\partial_{\alpha_j}u_n\rangle$ denotes the Berry connection. In the continuum limit, the EM are obtained by reducing the momentum $\mathbf{k}_c$ in Eq. \ref{eq:ger8} to the continuum momentum $\mathbf{k}$.
%In Eqs. (\ref{eq:ger8}), we drop the index $c$ from $\mathbf{r}_c$ and $\mathbf{k}_c$ to simplify the notation.

\subsection{ LDC and orbital magnetization } 
%First, we obtain the corresponding EM via wave packet dynamics. Secondly, using the EM,  We assume the velocity field varies slowly in position and time, allowing us to apply local and adiabatic approximations to study the system. 
After reviewing the necessary techniques and concepts in wave packet dynamics and EM. We now derive the local density of current (LDC) using the generic method proposed in Ref. \cite{dong2020berry}, and extract the orbital magnetization from the LDC. To obtain the LDC, we introduce an auxiliary coupling term $-\hat{\mathbf{J}}\cdot\mathbf{K}(\mathbf{r},t)$ ($\mathbf{K}(\mathbf{r},t)$ is the auxiliary momentum field and $\hat{\mathbf{J}}=\nabla_{\hat{\mathbf{k}}} H_0$ is the velocity or current operator) to the Hamiltonian. In fact, the LDC is obtained in the limit where $\mathbf{K}(\mathbf{r}, t) = 0$. 
%Moreover, due to the form of the velocity operator $\nabla_{\hat{\mathbf{k}}} H_0$, the auxiliary momentum field $\mathbf{K}(\mathbf{r}, t)$ can be absorbed into the momentum operator, i.e., $\hat{\mathbf{k}} \rightarrow \hat{\mathbf{k}} + \mathbf{K}$.
The total system is described by the following Hamiltonian

\begin{equation}
\label{eq:ger90}
H\left(\hat{\mathbf{k}},\mathbf{r};\mathbf{r},t\right) =H_0 \left(\hat{\mathbf{k}},\mathbf{r}\right)-\hat{\mathbf{k}}\cdot \mathbf{v} \left(\mathbf{r},t\right)-\hat{\mathbf{J}}\cdot\mathbf{K}(\mathbf{r},t),
\end{equation}
The set of slowly varying perturbations $\theta(\mathbf{r}, t)$ in equation (\ref{eq:ger1}) now takes the form ${\mathbf{K}(\mathbf{r}, t), \mathbf{v}(\mathbf{r}, t)}$. The zero order local Hamiltonian has the following form
\begin{equation}
\label{eq:ger9}
H_c=H_0 \left(\hat{\mathbf{k}},\mathbf{r}\right)-\hat{\mathbf{k}}\cdot \mathbf{v} \left(\mathbf{r}_c,t\right)-\hat{\mathbf{J}}\cdot\mathbf{K}(\mathbf{r}_c,t)\equiv \hat{H}.
\end{equation}
%We now derive the EM for a space-dependent velocity field and a time-dependent velocity field separately. On one hand, for a space-dependent velocity field $\mathbf{v}\left(\mathbf{r}\right)$, the wave packet is given by  \cite{23}:
In the derivation of the LDC, we need the EM for the Hamiltonian \ref{eq:ger90}. Therefore, we first obtain the EM by applying the semiclassical wave-packet dynamics method, as discussed in the previous subsection, to the Hamiltonian \ref{eq:ger90}.
The wave packet is constructed as
\begin{equation}
\label{eq:psi}
|\Psi\rangle=\int d^3\mathbf{k} a(\mathbf{k}) e^{i\mathbf{k}\cdot\mathbf{x}}|u_n\left(\mathbf{k},\mathbf{v},\mathbf{K},t\right)\rangle,
\end{equation}
where $\mathbf{k}$ is the eigenvalue of the momentum operator $\hat{\mathbf{k}}$, and $|u_n\left(\mathbf{k},\mathbf{v} \left(\mathbf{r}_c,t\right),\mathbf{K} \left(\mathbf{r}_c,t\right),t\right)\rangle$ is the eigenstate of the local Hamiltonian $H_c$, with the corresponding eigenvalue is $\epsilon_n \left(\mathbf{v} \left(\mathbf{r}_c,t\right),\mathbf{K}  \left(\mathbf{r}_c,t\right)\right)$.  The equations (\ref{eq:ger4.0}) and ( \ref{eq:ger4}) are transformed into

\begin{equation}
%\label{eq:ger4.0}
\langle\Psi | \hat{H} | \Psi\rangle = \epsilon_n \left(\mathbf{v} \left(\mathbf{r}_c,t\right),\mathbf{K}  \left(\mathbf{r}_c,t\right)\right),
\end{equation}
\begin{equation}
%\label{eq:ger4}
\delta\epsilon_n = \text{Im} \langle \partial_{k_i}u_n|\epsilon-\hat{H}|\partial_{\mathbf{v}}u_n\rangle\cdot\partial_{r_i}\mathbf{v}+\text{Im} \langle \partial_{k_i}u_n|\epsilon-\hat{H}|\partial_{\mathbf{K}}u_n\rangle\cdot\partial_{r_i}\mathbf{K},
\end{equation}
The dynamic part takes the same form as Eq. (\ref{eq:ger5}), and the Lagrangian reads
\begin{equation}
\label{eq:ger6.0}
L  =  \mathcal{A}_t+\dot{\mathbf{r}}_c\cdot\mathcal{A}_{\mathbf{r}_c}+\dot{\mathbf{k}_c}\cdot\mathcal{A}_{\mathbf{k}_c} +\dot{\mathbf{k}_c} \cdot \mathbf{r}_c- E,
\end{equation}
where the $\mathcal{A}_i=\langle u_n| i\partial_i |u_n\rangle$ for $i\in\{\mathbf{k}_c,\mathbf{r}_c,t\}$ are Berry connections, and the energy is given by
\begin{equation}
E=\epsilon+\delta\epsilon.
\end{equation}
To simplify the notation, we have omitted the center position label $c$ and the band index $n$, and will continue to do so henceforth. The EM is modified to
\begin{equation}
\label{eq:ger8.9}
\begin{cases}
\dot{\mathbf{r}}=\partial_{\mathbf{k}}E-\Omega_{\mathbf{k} t}-\Omega_{\mathbf{k} \mathbf{r}}\cdot \dot{\mathbf{r}}-\Omega_{\mathbf{k} \mathbf{k}}\cdot\dot{\mathbf{k}},\\ 
\dot{\mathbf{k}}= -\partial_{\mathbf{r}}E+\Omega_{\mathbf{r} t}+\Omega_{\mathbf{r} \mathbf{r}}\cdot\dot{\mathbf{r}}+ \Omega_{\mathbf{r}\mathbf{k}}\cdot\dot{\mathbf{k}}.
\end{cases}
\end{equation}

After deriving the necessary EM, we proceed to derive the LDC. The LDC is expressed as follows (see Eq. (10) in the supplementary material of Ref. \cite{dong2020berry}):
%From the LDC, we can extract the orbital magnetization. 
\begin{equation}
\label{eq:ger8.99}
\rho^{\mathbf{J}}_{\text{loc}} = \lim_{\mathbf{K}\rightarrow0}\left\{-\int d\mathbf{k} \, D f \left[\frac{\partial E}{\partial \mathbf{K}} - \Omega_{\mathbf{K}\mathcal{T}}\right] - \boldsymbol{\nabla}\cdot\int d\mathbf{k} \, D f \, \text{Im} \left\langle \partial_{\mathbf{k}} u_n \mid \epsilon - \hat{H} \mid \partial_{\mathbf{K}} u_n \right\rangle\right\}
\end{equation}
where $\Omega_{\mathbf{K}\mathcal{T}} = \dot{\mathbf{k}} \cdot \Omega_{\mathbf{Kk}} + \dot{\mathbf{r}} \cdot \Omega_{\mathbf{Kr}} + \Omega_{\mathbf{K}t}$, the local equilibrium distribution $f=\left(1+e^{\beta\epsilon\left(\mathbf{v,K}\right)}\right)^{-1}$, and $D \equiv 1 + \Omega_{k_i r_i}$ serves as the phase space measure. %It should be noted that the variable $\mathbf{h}$ in Ref. \cite{dong2020berry} corresponds to $-\mathbf{K}$ in our paper. 

In the second term of $\rho^{\mathbf{J}}_{\text{loc}}$, it suffices to set $D = 1$, as $\rho^{\mathbf{J}}_{\text{loc}}$ is considered only up to first order \cite{dong2020berry}. Additionally, the second term, being the divergence of a vector, clearly contributes to the orbital magnetization based on the definition of LDC in Ref. \cite{dong2020berry}. However, for the first term in $\rho^{\mathbf{J}}_{\text{loc}}$, further calculations are necessary to clarify how it contributes to the orbital magnetization. We now provide these details. 

Using the EM (\ref{eq:ger8.9}), the first term within the the $\{\}$ of $\rho^{\mathbf{J}}_{\text{loc}}$ is given by

\begin{eqnarray}
   &&-\int d\mathbf{k} D f \left[\frac{\partial E}{\partial\mathbf{K}}-\Omega_{\mathbf{K}\mathcal{T}}\right] \\
   &=-&\int d\mathbf{k} f \left[\frac{\partial E}{\partial\mathbf{K}}+\Omega_{\mathbf{K}k_i}\partial_{r_i}\epsilon - \Omega_{\mathbf{K}r_i}\partial_{k_i}\epsilon - \Omega_{\mathbf{K}t}+\Omega_{k_ir_i}\frac{\partial\epsilon}{\partial \mathbf{K}}\right] \nonumber\\
   &=-& \int d\mathbf{k} \left[\frac{\partial g}{\partial\mathbf{K}}- f \Omega_{\mathbf{K}t}+\frac{\partial (\Omega_{k_ir_i} g)}{\partial \mathbf{K}}+\frac{\partial (\Omega_{ \mathbf{K}k_i}g)}{\partial r_i}-g(\partial_{k_i}\Omega_{\mathbf{K}r_i}+\partial_{r_i}\Omega_{k_i \mathbf{K}}+\partial_{\mathbf{K}}\Omega_{r_ik_i })\right],
  \end{eqnarray} 
where the function $g=-\frac{1}{\beta}\ln ^{1+e^{-\beta\epsilon\left(\mathbf{v,K}\right)}}$ arises due to integration by parts. By applying the Bianchi identity $\partial_{k_i}\Omega_{\mathbf{K}r_i}+\partial_{r_i}\Omega_{k_i \mathbf{K}}+\partial_{\mathbf{K}}\Omega_{r_ik_i }=0$, the above equation can be simplified as follows:
\begin{eqnarray}\label{eq:array1}
   -\int d\mathbf{k} D f \left[\frac{\partial E}{\partial\mathbf{K}}-\Omega_{\mathbf{K}\mathcal{T}}\right] &=& -\int d\mathbf{k} \left[\frac{\partial g}{\partial\mathbf{K}}- f \Omega_{\mathbf{K}t}+\frac{\partial (\Omega_{k_ir_i} g)}{\partial \mathbf{K}}+\frac{\partial (\Omega_{ \mathbf{K}k_i}g)}{\partial r_i}\right]\nonumber\\
   &=-&  \partial_{\mathbf{K} }\int d\mathbf{k} (1+\Omega_{k_ir_i}) g-  \partial_{r_i} \int d\mathbf{k} \Omega_{ \mathbf{K}k_i}g+ \int d \mathbf{k} f \Omega_{\mathbf{K}t} \nonumber\\
     &=-&  \partial_{\mathbf{K} }\int d\mathbf{k} (1+\Omega_{k_ir_i}) g+\boldsymbol{\nabla} \cdot \int d\mathbf{k} \Omega_{ \mathbf{kK}}g+ \int d \mathbf{k} f \Omega_{\mathbf{K}t},
  \end{eqnarray} 
Since the vector potential $\mathbf{K}$ is minimally coupled to the momentum in the Hamiltonian through $\mathbf{k} - \mathbf{K}$, the derivative with respect to $\mathbf{K}$ can be replaced by the derivative with respect to $\mathbf{k}$. Note that the second term in the last equality of Eq. \ref{eq:array1} is a total divergence of a vector; hence it will contribute to the orbital magnetization. By combining the first and second terms of Eq. (\ref{eq:ger8.99}) and taking the limit $\mathbf{K}\rightarrow0$, we derive the LDC as:
\begin{eqnarray}
\label{eq:ger9.99}
   \rho^{\mathbf{J}}_{\text{loc}} &=&  -\partial_{\mathbf{k} }\int d\mathbf{k} (1+\Omega_{k_ir_i}) g-\boldsymbol{\nabla} \cdot \int d\mathbf{k} (-g\Omega_{ \mathbf{kk}}+f \text{Im} \langle \partial_{\mathbf{k}}u_n|\epsilon-\hat{H}|\partial_{\mathbf{k}}u_n\rangle)+\int d \mathbf{k} f \Omega_{\mathbf{k}t} \nonumber\\
     &=& -\boldsymbol{\nabla} \cdot \int d\mathbf{k} (-g\Omega_{ \mathbf{kk}}+f \text{Im} \langle \partial_{\mathbf{k}}u_n|\epsilon-\hat{H}|\partial_{\mathbf{k}}u_n\rangle)+ \int d \mathbf{k} f \Omega_{\mathbf{k}t} \equiv -\partial_i M^{i\mathbf{J}}  + \int d \mathbf{k} f \Omega_{\mathbf{k}t},
  \end{eqnarray}
where the second equality holds because $\mathbf{k}$ has already been integrated out.
%the first equality we use the fact that the derivative with respect to $\mathbf{K}$ will ultimately be replaced by the derivative with respect to $\mathbf{k}$, and

From the LDC (Eq. \ref{eq:ger9.99}), we can extract the orbital magnetization \cite{dong2020berry}
\begin{equation}
%\label{eq:ger9.91}
M^{iJ_j} = \int d\mathbf{k} (-g\Omega_{k_ik_j}+f \text{Im} \langle \partial_{k_i}u_n|\epsilon-\hat{H}|\partial_{k_j}u_n\rangle) \equiv M_{ij},
\end{equation}
Its vector form is given by
\begin{equation}
M^{\text{orb}}_{l}=\frac{\epsilon_{lij}M_{ij} }{2}= \int d\mathbf{k} \frac{1}{2}(-g\epsilon_{lij}\Omega_{k_ik_j}+f \epsilon_{lij} \text{Im} \langle \partial_{k_i}u_n|\epsilon-\hat{H}|\partial_{k_j}u_n\rangle) = \int d\mathbf{k} (f m^{\text{orb}}_l - g\Omega_{l}\left(\mathbf{k}\right)), 
\end{equation}
where $\Omega_{l}\left(\mathbf{k}\right)$ is the $l^{th}$ component of the Berry curvature, and $m_{l}^{\text{orb}}$ is the $l^{th}$ component of the orbital magnetic moment. Notice that after taking the limit $\mathbf{K} \rightarrow 0$, the distribution functions $f$ and $g$ take the following form 
\begin{equation}
\begin{cases}
f=\frac{1}{1+e^{\beta(\epsilon-\mathbf{k}\cdot\mathbf{v}})},\\
g=-\frac{1}{\beta}\ln[1+e^{-\beta(\epsilon-\mathbf{k}\cdot\mathbf{v})}].
\end{cases}
\end{equation}
Ultimately, by accounting for the band index and combining the effects from each band, we can derive the formula for the density of orbital magnetization. This accounts for the cumulative effect of all individual bands, providing a complete representation of the orbital magnetization density in the system, which is now given by:
 \begin{equation}
\mathbf{M}^{\text{orb}}= \sum_n\int d\mathbf{k} \{f \mathbf{m}_n^{\text{orb}} +\frac{1}{\beta}\ln[1+e^{-\beta(\epsilon_n-\mathbf{k}\cdot\mathbf{v})}]\boldsymbol{\Omega}_n\}, 
\end{equation}

Conversely, for a filled band, when using a gauge where $\mathcal{A}_t$ is periodic in the Brillouin zone, the last term in Eq. (\ref{eq:ger9.99}) can be written as:
\begin{equation}
\int d \mathbf{k} \Omega_{\mathbf{k}t} = -\partial_t \int d \mathbf{k} \mathcal{A_{\mathbf{k}}} = -\partial_t \mathbf{P},
\end{equation}
where $\mathbf{P}$ denotes the polarization. As a result, this final term represents the current generated by the polarization.}}

\section{Orbital magnetization of a parabolic band under a static velocity field}
\label{app:C}

In this Appendix, we derive the EM for a parabolic band with Berry phases due to internal degrees of freedom using the analogy between the velocity field and the electromagnetic vector potential. We then obtain the orbital magnetization by varying the appropriate equilibrium free energy density.
% a system driven by a velocity field that depends only on spatial coordinates. We also use the free energy density to derive the orbital magnetization.

\subsection{Alternate derivation of the EM}

Consider electrons described by a Bloch Hamiltonian $H_{0}\left(\mathbf{k}\right)$, where a band near its bottom can be approximated by a parabolic dispersion with an effective mass $m$. Upon applying an external static velocity field, the effective Hamiltonian is modified as:
\begin{equation}
\label{eq:ger9}
H\left(\mathbf{k},\mathbf{r}\right) = \frac{\mathbf{k}^2}{2m}-\mathbf{k}\cdot \mathbf{v} \left(\mathbf{r}\right).
\end{equation}
For sufficiently weak velocity fields compared to the typical group velocity, we can approximate the Hamiltonian as follows :
\begin{equation}
\label{eq:gern1}
H\approx\frac{ \left(\mathbf{k}-m\mathbf{v}\left(\mathbf{r}\right) \right)^2}{2m} .
\end{equation}

Noticing that this Hamiltonian is similar to the local Hamiltonian of electrons under magnetic fields in Refs. \cite{sundaram1999wave, xiao2010berry,23}, with $\mathbf{v}\left(\mathbf{r}_c\right)$ a similar role to the magnetic vector potential $\mathbf{A}\left(\mathbf{r}_c\right)$, and the angular velocity $\boldsymbol{\mathcal{V}}\equiv\frac{1}{2}\nabla\times\mathbf{v}$ can be identified as the magnetic field $\mathbf{B}=\nabla\times\mathbf{A}$.
Given the aforementioned similarities between the velocity $\mathbf{v}$ and the magnetic vector potential $\mathbf{A}$, and between the angular velocity $\boldsymbol{\mathcal{V}}$ and the magnetic field $\mathbf{B}$, we immediately obtain the Lagrangian for our model as:
\begin{equation}
\label{eq:lagg1}
L=\mathbf{k}\cdot\dot{\mathbf{r}}-\dot{\mathbf{r}}\cdot m\mathbf{v}+\dot{\mathbf{k}}\cdot\mathcal{A}\left(\mathbf{k}\right)-h\left(\mathbf{k}\right),
\end{equation}
with the energy $h\left(\mathbf{k}\right)=\epsilon_n\left(\mathbf{k}\right)-2m\boldsymbol{\mathcal{V}}\cdot\mathbf{m}^{\text{orb}}_n$. The Eq. (\ref{eq:lagg1}) is similar to the Eq. (3.7) in Ref. \cite{sundaram1999wave}. Plugging the Eq. (\ref{eq:lagg1}) into the Euler-Lagrange equations, one can obtain the EM as

\begin{equation}
\label{eq:em}
\begin{cases}
\dot{\mathbf{r}}=& \partial_{\mathbf{k}}h-\dot{\mathbf{k}}\times\boldsymbol{\mathbf{\Omega}}_{n} ,\\
\dot{\mathbf{k}}= & -\partial_{\mathbf{r}}h-\dot{\mathbf{r}}\times2m\boldsymbol{\mathcal{V}},
\end{cases}
\end{equation}
where $\boldsymbol{\mathbf{\Omega}}_{n}$ denotes the Berry curvature of the $n$th band. As expected, the above EM is similar to the EM for an electron under a magnetic field, such as Equation (3.8) in Ref. \cite{sundaram1999wave} and Equations (5.8a,b) in Ref. \cite{xiao2010berry}. 

The Berry curvature in above EM implies that the $\mathbf{r,k}$ are not canonical coordinates. In reminder of this section, we investigate the impact of the equations of motion on the phase space spanned by noncanonical coordinates. In canonical coordinates, denoted as $\eta = (\mathbf{q}, \mathbf{p})$, the Hamilton equations can be expressed as $\dot{\eta}^{\alpha}\theta_{\alpha\beta} = \partial_{\beta}h$, where the antisymmetric matrix  $\theta\equiv J=(0,1;-1,0)$ is known as the symplectic form\cite{kamenev2023field}. This establishes the foundation for understanding the dynamics in canonical coordinates.

For the equations of motion presented in Eq. (\ref{eq:em}), the corresponding symplectic form exhibits a distinct structure. It can be expressed as:

\begin{equation}
\theta_{\alpha\beta}=
\begin{pmatrix}
&\epsilon_{ijl}2m\boldsymbol{\mathcal{V}}^{l} &\delta_{ij} \\
& -\delta_{ij}&-\epsilon_{ijl}\Omega_n^{l} \\
\end{pmatrix}
\end{equation}
Here, $\alpha$ and $\beta$ represent elements from the set $\{\mathbf{r}, \mathbf{k}\}$, while $i$, $j$, and $l$ take values from the set $\{x, y, z\}$. This modified symplectic form reveals new insights into the equations of motion in noncanonical coordinates. It highlights the influence of the antisymmetric matrix and the additional terms that arise due to the distinct structure of the symplectic form.

In the canonical coordinates $\eta = (\mathbf{q}, \mathbf{p})$, the phase-space volume element is given by $dV = d\mathbf{q}d\mathbf{p}$. However, when changing coordinates to noncanonical coordinates $\eta \rightarrow \zeta = (\mathbf{r}, \mathbf{k})$, the symplectic form undergoes a transformation. Specifically, $J_{\alpha\beta} \rightarrow \theta_{\alpha\beta} = \frac{\partial \eta^\sigma}{\partial\zeta^{\alpha}}\frac{\partial \eta^\gamma}{\partial\zeta^{\beta}}J_{\sigma\gamma}$. This transformation of the symplectic form leads to a corresponding transformation in the phase-space volume element.

The volume element in noncanonical coordinates is given by $dV = \sqrt{|\det{\theta}|}d\mathbf{r}d\mathbf{k} = (1+2m\boldsymbol{\Omega}_{n}\cdot{\boldsymbol{\mathcal{V}}})d\mathbf{r}d\mathbf{k}$. This expression elucidates the modification to the volume element due to the transformation and emphasizes the role of the additional terms involving the parameters $\boldsymbol{\Omega}_{n}$ and $\boldsymbol{\mathcal{V}}$.

These results shed light on the structure of the symplectic form in noncanonical coordinates and its impact on the phase-space volume element. Understanding these transformations is vital for comprehending the dynamics and exploring various physical systems with different coordinate choices.

\subsection{Free energy and orbital magnetization}

In this part, we employ the modified phase-space volume, which in turn modifies the free energy density, to calculate the orbital magnetization response to the velocity field to linear order. 

Consider a probability distribution function over the phase space volume, denoted as $n(\mathbf{r,k,t})(1+2\boldsymbol{\Omega}_n\cdot m\boldsymbol{\mathcal{V}})d\mathbf{r}d\mathbf{k}$. Under a Hamiltonian flow (without collisions), it evolves according to
\begin{equation}
\label{eq:F0}
\frac{\partial n}{\partial t}+\partial_{\mathbf{r}}\left(n\dot{\mathbf{r}}\right)+\partial_{\mathbf{k}}\left(n\dot{\mathbf{k}}\right)=-n\frac{d_{t}\left(2\boldsymbol{\Omega}_n\cdot m\boldsymbol{\mathcal{V}}\right)}{1+2\boldsymbol{\Omega}_n\cdot m\boldsymbol{\mathcal{V}}},
\end{equation}
where $d_{t}\left(2\boldsymbol{\Omega}_n\cdot m\boldsymbol{\mathcal{V}}\right)\equiv\partial_{t}\left(2\boldsymbol{\Omega}_n\cdot m\boldsymbol{\mathcal{V}}\right)+\partial_{\mathbf{r}}\left(2\boldsymbol{\Omega}_n\cdot m\boldsymbol{\mathcal{V}}\right)\cdot\dot{\mathbf{r}}+\partial_{\mathbf{k}}\left(2\boldsymbol{\Omega}_n\cdot m\boldsymbol{\mathcal{V}}\right)\cdot\dot{\mathbf{k}}$. 
The Eq. (\ref{eq:F0}) does not have a form of the continuity relation, due to the presence of the right-hand side. This reflects the fact that $\int d\mathbf{r}d\mathbf{k}n\left(\mathbf{r,k},t\right)$ is not conserved. However, the quantity $\int d\mathbf{r}d\mathbf{k}\tilde{n}\left(\mathbf{r,k},t\right)\equiv\int d\mathbf{r}d\mathbf{k}n\left(\mathbf{r,k},t\right)(1+2\boldsymbol{\Omega}_n\cdot m\boldsymbol{\mathcal{V}})$ remains conserved. Therefore, any observables should be expressed as $O_{t}=\int d\mathbf{r}d\mathbf{k}(1+2\boldsymbol{\Omega}_n\cdot m\boldsymbol{\mathcal{V}})n(\mathbf{r,k},t)O(\mathbf{r,k},t)$. Also, $\tilde{n}(\mathbf{r,k},t)\equiv(1+2\boldsymbol{\Omega}_n\cdot m\boldsymbol{\mathcal{V}})n(\mathbf{r,k},t)$ satisfies the continuity equation: 

\begin{equation}
\label{eq:F1}
\frac{\partial \tilde{n}}{\partial t}+\dot{\mathbf{r}}\cdot\partial_{\mathbf{r}}\tilde{n}+\dot{\mathbf{k}}\cdot\partial_{\mathbf{k}}\tilde{n}=0. 
\end{equation}
Similar to $\mathbf{B}$ in the chiral kinetic theory, $\boldsymbol{\mathcal{V}}$ modifies the phase space measure in the free energy density:
\begin{align}
\label{eq:F}
F = -\frac{1}{\beta}\sum_{n}\intop_{\mathbf{k}}(1+2\mathbf{\boldsymbol{\Omega}}_{n}(\mathbf{k})\cdot m{\boldsymbol{\mathcal{V}}})\ln\left(1+e^{-\beta\left({\epsilon_{n,\mathbf{k}}-\mathbf{k}\cdot \mathbf{v}-2\mathbf{m}^\text{orb}_{n}(\mathbf{k})}\cdot m\boldsymbol{\mathcal{V}}\right)}\right).
\end{align}

%The derivation of $\mathfrak{M}_n$ and the measure in Eq. (\ref{eq:F}) for the free energy density represents a significant outcome of this study.
The converse vortical effect refers to the response of orbital magnetization to velocity. In order to calculate the density of orbital magnetization, we differentiate the free energy density with respect to $2 m\boldsymbol{\mathcal{V}}$ while keeping the temperature $T=\beta^{-1}$ fixed. This calculation leads to:
\begin{align}
\mathbf{M}^\text{orb}(\mathbf{v}) &= -\frac{\delta F}{\delta{(2 m\boldsymbol{\mathcal{V}}})}_{|\boldsymbol{\mathcal{V}}=0} \nonumber \\
&= \sum_{n}{\int_{\mathbf{k}}\mathbf{m}^\text{orb}_{n}(\mathbf{k}) f (\epsilon_{n,\mathbf{k}},\mathbf{v})} \nonumber \\
&\quad + \frac{1}{\beta}\int_{\mathbf{k}}\boldsymbol{\Omega}_{n}(\mathbf{k})\ln\left(1+e^{-\beta\left(\epsilon_{n,\mathbf{k}}-\mathbf{k\cdot v}\right)}\right) \nonumber \\
&\equiv{ \chi^\text{orb}{\cdot \mathbf{v}}}+O\left(\mathbf{v^{2}}\right)
\end{align}
where Fermi distribution function \cite{shitade2020chiral} ${f\left(\mathrm{\epsilon_{n,\mathbf{k}}},\mathbf{v}\right)}\equiv\left(e^{\beta\left(\mathrm{\epsilon_{n,\mathbf{k}}}-\mathbf{k\cdot v}\right)}+1\right)^{-1}$, and the tensor $\chi^\text{orb}$ representing the orbital magnetic susceptibility is denoted as:
\begin{align}
\label{eq:chi_stat0}
\chi_{ij}^\text{orb} &= -{\sum_{n}\int_{\mathbf{k}} m_{n,i}^\text{orb}(\mathbf{k})f^{'}(\epsilon_{n,\mathbf{k}})k_{j}} \nonumber \\
&\quad + \sum_{n}\int_{\mathbf{k}}f(\epsilon_{n,\mathbf{k}})\Omega_{n,i}(\mathbf{k})k_{j}\equiv \chi_{ij}^\text{Fs} +\chi_{ij}^\text{occ}
\end{align}
At $T=0$, equation (\ref{eq:chi_stat0}) coincides with Eq. (\ref{eq:chi_stat}) in the main text. The Eq. (\ref{eq:chi_stat0}) reveals that the magnetic susceptibility is determined by the orbital magnetic moment of electrons on the Fermi surface (indicated as $\chi_{ij}^\text{Fs} $) as well as the Berry curvature of the occupied bands (indicated as $\chi_{ij}^\text{occ}$).

\section{ Kubo formla for orbital magnetic susceptibility} %%#######
\label{app:D}
In this section, we present a derivation of the orbital magnetic susceptibility for a clean (disordered) electron fluid in both the static and uniform limit. The calculation is similar to the one that yields the vortical effect\cite{nanda2023vortical}, which we refer the reader to for a more detailed description. 

\subsection{Orbital magnetic susceptibility as a functional of Green's function and current operator}

In the continuum, the perturbation induced by the velocity field can be written as: $H_1\equiv -i\mathbf{v}\left(\mathbf{r},t\right)\cdot\nabla_{\mathbf{r}}$. In the Bloch basis, the perturbation matrix is composed of $\left<u_{m,\bk}|\bk|u_{n,\bk+\bq}\right>\cdot\mathbf{v}(\bq,t)$ and $\left<u_{m,\bk}|(-i\boldsymbol{\nabla}_{\brho})|u_{n,\bk+\bq}\right>\cdot\mathbf{v}(\bq,t)$. The first term arises from the plane wave component of the Bloch function $\psi^{n}_{\bk}(\br)$, while the second term is a result of the periodic part of the Bloch function. The details are as follows:

The Bloch wavefunction for the $n$-th band is generically of the form $\psi^{n}_{\bk}(\br)\equiv\psi^{n}_{\bk}(\mathbf{R}+\brho)=N^{-1/2}e^{i\bk\cdot(\mathbf{R}+\brho)}u_{n,\bk}(\brho)$, where $N$ is the number of unit cells, $\mathbf{R}$ is a discrete index that labels them, $\brho$ denotes position within a unit cell, and $u_{n,\bk}(\brho)$ is periodic in $\brho$ with the same periodicity as the underlying Hamiltonian.
In this basis,
\begin{eqnarray}
   \langle\psi^{m}_{\bk}|-i\mathbf{v}\left(\mathbf{r},t\right)\cdot\nabla_{\mathbf{r}}|\psi^{n}_{\bk+\bq'}\rangle =-i\intop_{\br} \psi^{m*}_{\bk}(\br)\mathbf{v}\left(\mathbf{r},t\right)\cdot\nabla_{\mathbf{r}} \psi^{n}_{\bk+\bq'}(\br)
\end{eqnarray}
Assuming $\mathbf{v}\left(\mathbf{r},t\right)=e^{-i\bq\cdot\br}\mathbf{v}\left(\mathbf{q},t\right)$ and approximating $\br\sim\mathbf{R}$ and  $\boldsymbol{\nabla}_{\br}=\boldsymbol{\nabla}_{\brho}$, the matrix element becomes 

\begin{eqnarray}
   \langle\psi^{m}_{\bk}|-i\mathbf{v}\left(\mathbf{r},t\right)\cdot\nabla_{\mathbf{r}}|\psi^{n}_{\bk+\bq'}\rangle &=& -\frac{1}{N}\sum_{\mathbf{R}}e^{i(\bq'-\bq)\cdot\mathbf{R}} \intop_{\brho,\brho'}e^{-i\bk\cdot\brho+i(\bk+\bq'-\bq)\cdot\brho'}\left[u^{*}_{m,\bk}(\brho) i\boldsymbol{\nabla}_{\brho}\delta\left(\brho-\brho' \right)u_{n,\bk+\bq'}(\brho')\right]\cdot\mathbf{v}(\bq,t) \nonumber\\
   &=& \frac{(2\pi)^3}{N}\sum_{\bK}\delta(\bq'-\bq-\bK)\left[ \intop_{\brho}e^{i(\bq'-\bq)\cdot\brho}u^{*}_{m,\bk}(\brho)\left(\mathbf{k}-i\boldsymbol{\nabla}_{\brho}\right)u_{n,\bk+\bq'}(\brho)\right]\cdot\mathbf{v}(\bq,t)\nonumber\\
   &=&  \frac{(2\pi)^3}{N}\sum_{\bK}\delta(\bq'-\bq-\bK)\left[ \intop_{\brho}e^{i\bK\cdot\brho}u^{*}_{m,\bk}(\brho)\left(\mathbf{k}-i\boldsymbol{\nabla}_{\brho}\right)u_{n,\bk+\bq+\bK}(\brho)\right]\cdot\mathbf{v}(\bq,t)
  \end{eqnarray}
where $\bK$ are reciprocal lattice vectors. Since $u_{n,\bk+\bq+\bK}(\brho)=e^{-i\bK\cdot\brho}u_{n,\bk+\bq}(\brho)$, the equation above can be expressed as follows:
\begin{eqnarray}
   &&\langle\psi^{m}_{\bk}|-i\mathbf{v}\left(\mathbf{r},t\right)\cdot\nabla_{\mathbf{r}}|\psi^{n}_{\bk+\bq'}\rangle = (2\pi)^3\delta(\bq'-\bq)\left[ \intop_{\brho}u^{*}_{m,\bk}(\brho)\left(\mathbf{k}-i\boldsymbol{\nabla}_{\brho}\right)u_{n,\bk+\bq}(\brho)\right]\cdot\mathbf{v}(\bq,t)
  \end{eqnarray}
Each term in the sum over $\bK$ gives the same contribution and cancels the factor of $N$. Thus, we can safely assume $\bq$ and $\bq'$ to be within the first Brillouin zone and write
\begin{eqnarray}
    \langle\psi^{m}_{\bk}|-i\mathbf{v}\left(\mathbf{r},t\right)\cdot\nabla_{\mathbf{r}}|\psi^{n}_{\bk+\bq}\rangle = (2\pi)^3\left<u_{m,\bk}|(\bk-i\boldsymbol{\nabla}_{\brho})|u_{n,\bk+\bq}\right>\cdot\mathbf{v}(\bq,t)\equiv (2\pi)^3\left<u_{m,\bk}|\hat{\mathbf{Q}}|u_{n,\bk+\bq}\right>\cdot\mathbf{v}(\bq,t)
\end{eqnarray}
These are the matrix elements of the perturbation in the Bloch basis, and they enter into Kubo's formula, giving the orbital magnetization response to an external velocity field.

The orbital magnetization response to an external velocity field is captured by the response function, which can be expressed as a function of the Green's function and current operator :
\begin{align}
\label{eq:d5}
\chi_{ij}^{\text{orb}}(\mathbf{q},iq_{n}) &= -\frac{1}{2}\epsilon_{i\mu\nu}i\partial_{q_{\mu}}T\sum_{i\nu_{n}} \intop_{\mathbf{k}}\text{tr}\left[J_{\nu}(\mathbf{k}+\mathbf{q})G_{0}(\mathbf{k},i\nu_{n})\right.\hat{Q}_{j} \left. G_{0}(\mathbf{k}+\mathbf{q},i\nu_{n}+iq_{n})\right],
\end{align}
Here, $G_{0}(\mathbf{k},i\nu_{n})$ represents the standard unperturbed Matsubara Green's function, defined as $\left[i\nu_{n}-H_{0}(\mathbf{k})+i \text{sgn}(\nu_n)/2\tau\right]^{-1}$, where $H_{0}(\mathbf{k})$ denotes the unperturbed Hamiltonian. The $\nu$-th component of the current operator is denoted as $J_{v}$ and is given by $\frac{\partial H_{0}}{\partial k_{\nu}}$. Furthermore, we introduce $\hat{\mathbf{Q}}\equiv \hat{\mathbf{k}}-i\nabla_{\boldsymbol{\rho}}$, where $i{\nabla}_{\boldsymbol{\rho}}$ represents the modification arising from the lattice background\cite{nanda2023vortical}. In the continuum limit, $\hat{\mathbf{Q}}$ converges to $\hat{\mathbf{k}}$.

\subsection{Magnetic susceptibility under different limits}
After establishing the connection between the magnetic susceptibility and the Green's function, we will provide detailed steps for deriving the final expression of the magnetic susceptibility and analyze its behavior under different limits.
Substituting the expression of $G_{0}=\left[i\nu_{n}-H_{0}(\mathbf{k})+i \text{sgn}(\nu_n)/2\tau\right]^{-1}$ into the Eq. (\ref{eq:d5}) for the magnetic susceptibility and defining $\mathbf{Q}^{mn}\equiv \langle u_{m,\mathbf{k}} |\hat{\mathbf{Q}}| u_{n,\mathbf{k}+\mathbf{q}}\rangle$, we obtain:
%\begin{widetext}
\begin{align}
\chi_{ij}^{\text{orb}}(\mathbf{q},iq_{n}) = &-\frac{1}{2}\epsilon_{i\mu\nu}i\partial_{q_{\mu}}T\sum_{i\nu_{n}}\intop_{\mathbf{k}}\sum_{n,m}\frac{\left\langle u_{n,\mathbf{k}+\mathbf{q}}\left|J_{\nu}(\mathbf{k}+\mathbf{q})\right|u_{m,\mathbf{k}}\right\rangle}{\left(i\nu_{n}-\epsilon_{m,\mathbf{k}}+i\frac{\text{sgn}(\nu_{n})}{2\tau}\right)} \frac{Q^{mn}_{j}}{\left(i\nu_{n}+iq_{n}+\epsilon_{n,\mathbf{k}+\mathbf{q}}+i\frac{\text{sgn}(\nu_{n}+q_{n})}{2\tau}\right)},
\end{align}
%\end{widetext}
which can be simplified as :
\begin{align}
\label{eq:chi1}
\chi_{ij}^{\text{orb}}(\mathbf{q},\omega) &= -\frac{i}{2}\epsilon_{i\mu\nu}\intop_{\mathbf{k}}\sum_{n,m}\left[\partial_{q_{\mu}}S_{m,n}\left(\mathbf{k},\mathbf{q},iq_{n}\right)\right] \left\langle u_{n,\mathbf{k}+\mathbf{q}}\left|J_{\nu}(\mathbf{k}+\mathbf{q})\right|u_{m,\mathbf{k}}\right\rangle Q^{mn}_{j} \nonumber \\
&\quad -\frac{i}{2}\epsilon_{i\mu\nu}\intop_{\mathbf{k}}\sum_{n,m} S_{m,n}\left(\mathbf{k},\mathbf{q},iq_{n}\right)\left[\partial_{q_{\mu}} \left\langle u_{n,\mathbf{k}+\mathbf{q}}\left|J_{\nu}(\mathbf{k}+\mathbf{q})\right|u_{m,\mathbf{k}}\right\rangle Q^{mn}_{j}\right],
\end{align}
where $|u_{n,\mathbf{k}}\rangle$ and $\epsilon_{n,\mathbf{k}}$ are the Bloch
eigenfunction and eigenenergy of the band $n$, respectively, and
the factor
%\begin{widetext}
\begin{align}
S_{m,n}\left(\mathbf{k},\mathbf{q},iq_{n}\right) &= T\sum_{i\nu_{n}}\frac{1}{\left(i\nu_{n}-\epsilon_{m,\mathbf{k}}+i\frac{\text{sgn}\left(\nu_{n}\right)}{2\tau}\right)} \frac{1}{\left(i\nu_{n}+iq_{n}-\epsilon_{n,\mathbf{k}+\mathbf{q}}+i\frac{\text{sgn}\left(\nu_{n}+q_{n}\right)}{2\tau}\right)},
\end{align}
%\end{widetext}
Performing the Matsubara summation and the analytical continuum $iq_{n}\rightarrow\omega+i0^{+}$,
we get
%\begin{widetext}
\begin{align}
S_{m,n}\left(\mathbf{k},\mathbf{q},\omega\right) &= -\int dz \text{Im}\left[\frac{2}{z+\frac{i}{2\tau}}\right] \frac{f\left(\epsilon_{m,\mathbf{k}}+z\right)-f\left(\epsilon_{n,\mathbf{k}+\mathbf{q}}-z\right)}{z+\epsilon_{m,\mathbf{k}}-\epsilon_{n,\mathbf{k}+\mathbf{q}}+\omega+\frac{i}{2\tau}},
\end{align}
%\end{widetext}
At the limit $\left(\mathbf{q},\omega\right)\rightarrow (\mathbf{0},0)$, we get:
\begin{align}
\label{eq:chi2}
\chi_{ij}^{\text{orb}}(\mathbf{0},0) &= -\frac{i}{2}\epsilon_{i\mu\nu}\intop_{\mathbf{k}}\sum_{n,m}\left[\frac{d S_{m,n}\left(\mathbf{k},\mathbf{0},0\right)}{d\epsilon_{n,\mathbf{k}}}\partial_{\mu}\epsilon_{n,\mathbf{k}}\right] \left\langle u_{n,\mathbf{k}}\left|\partial_{\nu}H_{0}(\mathbf{k})\right|u_{m,\mathbf{k}}\right\rangle Q^{mn}_{j} \nonumber \\
&\quad -\frac{i}{2}\epsilon_{i\mu\nu}\intop_{\mathbf{k}}\sum_{n,m} S_{m,n}\left(\mathbf{k},\mathbf{0},0\right)\left[ \left\langle \partial_{\mu}u_{n,\mathbf{k}}\left|\partial_{\nu}H_0(\mathbf{k})\right|u_{m,\mathbf{k}}\right\rangle Q^{mn}_{j}+\left\langle u_{n,\mathbf{k}}\left|\partial_{\nu}H_0(\mathbf{k})\right|u_{m,\mathbf{k}}\right\rangle\langle u_{m,\mathbf{k}}\mid \hat{Q}_{j}\mid\partial_{\mu}u_{n,\mathbf{k}}\rangle\right].
\end{align}
Note that the expression of $S_{m,n}\left(\mathbf{k},\mathbf{0},0\right)$ depends on the order of limits, which we will clarify shortly. At zero temperature and to leading order in $\tau^{-1}$, the factor $S_{m,n}\left(\mathbf{k},\mathbf{0},0\right)=\frac{\Theta\left(-\epsilon_{m,\mathbf{k}}\right)-\Theta\left(-\epsilon_{n,\mathbf{k}}\right)}{\epsilon_{m,\mathbf{k}}-\epsilon_{n\mathbf{k}}}$ for $m\neq n$, and  $S_{n,n}\left(\mathbf{k},\mathbf{0},0\right)=-\delta(\epsilon_{n,\mathbf{k}}) \quad \text{or } 0$. 
Using the relations
\begin{align}
\left\langle u_{n,\mathbf{k}}\left|\partial_{v}H_{0}(\mathbf{k})\right|u_{m,\mathbf{k}}\right\rangle &= -\left(\epsilon_{n,\mathbf{k}}-\epsilon_{m,\mathbf{k}}\right)\left\langle u_{n,\mathbf{k}}\mid\partial_{v}u_{m,\mathbf{k}}\right\rangle+ \delta_{n,m}\partial_{v}\epsilon_{n,\mathbf{k}},
\end{align}
and
\begin{align}
&\left\langle u_{n,\mathbf{k}}\left|\partial_{v}H_{0}(\mathbf{k})\right|\partial_{\theta}u_{m,\mathbf{k}}\right\rangle = \epsilon_{n,\mathbf{k}}\left\langle \partial_{v}u_{n,\mathbf{k}}\mid\partial_{\theta}u_{m,\mathbf{k}}\right\rangle -\left\langle \partial_{v}u_{n,\mathbf{k}}\mid H_{0}\left(\mathbf{k}\right)\mid\partial_{\theta}u_{m,\mathbf{k}}\right\rangle+\partial_{v}\epsilon_{n,\mathbf{k}}\left\langle u_{n,\mathbf{k}}\mid\partial_{\theta}u_{m,\mathbf{k}}\right\rangle.
\end{align}

The first term on the right-hand side of Eq. (\ref{eq:chi2}) can be expressed as follows:
\begin{align}
\label{eq:chi3}
 -\frac{i}{2}\epsilon_{i\mu\nu}\intop_{\mathbf{k}}\sum_{n\neq m}
 [\delta(\epsilon_{m,\mathbf{k}})+S_{m,n}\left(\mathbf{k},\mathbf{0},0\right)]\langle\partial_{\mu}u_{m,\mathbf{k}}\mid\partial_{\nu}\epsilon_{m,\mathbf{k}}\mid u_{n,\mathbf{k}}\rangle Q^{nm}_j,
\end{align}
and the second term on the right-hand side of Eq. (\ref{eq:chi2}) can be expressed as follows:
\begin{align}
\label{eq:chi4}
 &-\frac{i}{2}\epsilon_{i\mu\nu}\intop_{\mathbf{k}}\sum_{n\neq m}S_{m,n}\left(\mathbf{k},\mathbf{0},0\right)\left[\langle\partial_{\mu} u_{m,\mathbf{k}}\mid\partial_{\nu}H_{0}(\mathbf{k})\mid u_{n,\mathbf{k}}\rangle Q^{nm}_j +\langle u_{m,\mathbf{k}}\mid \partial_{\nu}H_0(\mathbf{k})\mid u_{n,\mathbf{k}}\rangle\langle u_{n,\mathbf{k}}\mid \hat{Q}_j\mid\partial_{\mu}u_{m,\mathbf{k}}\rangle\right] \nonumber \\
 & -\frac{i}{2}\epsilon_{i\mu\nu}\intop_{\mathbf{k}}\sum_{n}S_{n,n}\left(\mathbf{k},\mathbf{0},0\right)[\langle\partial_{\mu} u_{n,\mathbf{k}}\mid (\epsilon_{n,\mathbf{k}}-H_{0}(\mathbf{k}))\mid \partial_{\nu}u_{n,\mathbf{k}}\rangle Q^{nn}_j] \nonumber \\
 &-\frac{i}{2}\epsilon_{i\mu\nu}\intop_{\mathbf{k}}\sum_{n}S_{n,n}\left(\mathbf{k},\mathbf{0},0\right)\partial_{\nu}\epsilon_{n,\mathbf{k}}[\langle u_{n,\mathbf{k}}\mid\hat{Q}_j\mid\partial_{\mu}u_{n,\mathbf{k}}\rangle+\langle\partial_{\mu} u_{n,\mathbf{k}}\mid u_{n,\mathbf{k}}\rangle Q^{nn}_j],
\end{align}
Combining Eq. (\ref{eq:chi3}) and Eq. (\ref{eq:chi4}), finally, we obtain the expression for the magnetic susceptibility as:
%\begin{widetext}
\begin{align}
\label{eq:33}
\chi_{ij}^{\text{orb}}(\mathbf{0},0) &= -\intop_{\mathbf{k}}\sum_{n\neq m}S_{m,n}\left(\mathbf{k},\mathbf{0},0\right)\mathcal{M}^{mn}_{i}(\mathbf{k})Q^{nm}_j \nonumber \\
& +\frac{1}{2}\epsilon_{i\mu\nu}\intop_{\mathbf{k}}\sum_{n\neq m}S_{m, n}\left(\mathbf{k},\mathbf{0},0\right)(\epsilon_{m,\mathbf{k}}-\epsilon_{n,\mathbf{k}})\langle u_{m,\mathbf{k}}\mid \partial_{\nu}u_{n,\mathbf{k}}\rangle\langle u_{n,\mathbf{k}}\mid \partial_{\rho_j}\mid\partial_{\mu}u_{m,\mathbf{k}}\rangle  \nonumber \\
& + \frac{i}{2}\epsilon_{i\mu\nu}\intop_{\mathbf{k}}\sum_{n\neq m}\left[S_{m,n}\left(\mathbf{k},\mathbf{0},0\right)\left(\epsilon_{m,\mathbf{k}}-\epsilon_{n,\mathbf{k}}\right)\right] \left[A_{\nu}^{nm}\left(\mathbf{k}\right)A_{\mu}^{mn}\left(\mathbf{k}\right)\right]k_{j} \nonumber \\
 &-\frac{i}{2}\epsilon_{i\mu\nu}\intop_{\mathbf{k}}\sum_{n}S_{n,n}\left(\mathbf{k},\mathbf{0},0\right)[\langle\partial_{\mu} u_{n,\mathbf{k}}\mid (\epsilon_{n,\mathbf{k}}-H_{0}(\mathbf{k}))\mid \partial_{\nu}u_{n,\mathbf{k}}\rangle Q^{nn}_j] \nonumber \\
 & -\frac{i}{2}\epsilon_{i\mu\nu}\intop_{\mathbf{k}}\sum_{n\neq m}
 \delta(\epsilon_{m,\mathbf{k}})\langle\partial_{\mu}u_{m,\mathbf{k}}\mid\partial_{\nu}\epsilon_{m,\mathbf{k}}\mid u_{n,\mathbf{k}}\rangle Q^{nm}_j \nonumber \\
 &-\frac{i}{2}\epsilon_{i\mu\nu}\intop_{\mathbf{k}}\sum_{n}S_{n,n}\left(\mathbf{k},\mathbf{0},0\right)\partial_{\nu}\epsilon_{n,\mathbf{k}}[\langle u_{n,\mathbf{k}}\mid\hat{Q}_j\mid\partial_{\mu}u_{n,\mathbf{k}}\rangle+\langle\partial_{\mu} u_{n,\mathbf{k}}\mid u_{n,\mathbf{k}}\rangle Q^{nn}_j].
\end{align}
%&  - \sum_{n}\frac{i}{2}\intop_{\mathbf{k}}S_{n,n}\left(\mathbf{k},\mathbf{0},0\right)\left[\left\langle \nabla_\mathbf{k} u_{n,\mathbf{k}}\mid\times\left(\epsilon_{n,\mathbf{k}}-H_0(\mathbf{k})\right)\mid\nabla_\mathbf{k} u_{n,\mathbf{k}}\right\rangle \right]_{i} Q^{nn}_{j}\nonumber \\
where $\mathcal{M}^{mn}_{i}$ represents the inter-band orbital magnetization matrix for the Bloch electrons \cite{ogata2017theory}, which takes the following form,
\begin{align}
\mathcal{M}^{mn}_{i}(\mathbf{k})=\frac{i}{2}\epsilon_{i\mu\nu}\left[ \langle \partial_{\mu}u_{m,\mathbf{k}}\mid \left(\partial_{\nu}H_0(\mathbf{k})+\partial_{\nu}\epsilon_{m,\mathbf{k}}\right)\mid u_{n,\mathbf{k}}\rangle\right].
\end{align}

In some limit conditions, such as the nearly-free electron and deep tight-binding limits \cite{nanda2023vortical}, the term $\langle u_{n,\mathbf{k}}\mid\partial_{\rho_j} u_{m,\mathbf{k}}\rangle$ is negligible, and Eq. (\ref{eq:33}) can be further simplified as:
\begin{align}
\label{eq:35}
\chi_{ij}^{\text{orb}}(\mathbf{0},0) &=- \sum_{n}\frac{i}{2}\intop_{\mathbf{k}}S_{n,n}\left(\mathbf{k},\mathbf{0},0\right)\left[\left\langle \nabla_\mathbf{k} u_{n,\mathbf{k}}\mid\times\left(\epsilon_{n,\mathbf{k}}-H_0(\mathbf{k})\right)\mid\nabla_\mathbf{k} u_{n,\mathbf{k}}\right\rangle \right]_{i} k_{j} \nonumber \\
&\quad + \frac{i}{2}\epsilon_{i\mu\nu}\intop_{\mathbf{k}}\sum_{n\neq m}\left[S_{m,n}\left(\mathbf{k},\mathbf{0},0\right)\left(\epsilon_{m,\mathbf{k}}-\epsilon_{n,\mathbf{k}}\right)\right] \left[A_{\nu}^{nm}\left(\mathbf{k}\right)A_{\mu}^{mn}\left(\mathbf{k}\right)k_{j}\right].
\end{align}
%The contribution of the first term in the Eq. (\ref{eq:33}) vanishes at $q=0$ for a regularized $\mathbf{k}$ space since its integrand becomes a total derivative owing to $S_{n,n}(\mathbf{k})$ depending on $\mathbf{k}$ only through $\epsilon_{n,\mathbf{k}}$. 
The first term accounts for the contribution of the intra-band orbital magnetic moment to the orbital magnetization, while the second term reflects the dependence of the orbital magnetization on the Berry connection of the occupied bands. 
In the subsequent analysis, we assume the term $\langle u_{n,\mathbf{k}}\mid\partial_{\rho_j} u_{n,\mathbf{k}}\rangle$ is negligible and thoroughly examine the magnetic susceptibility under different limits, with a specific emphasis on the zero-temperature.

\subsubsection{Static limit ($\omega\rightarrow0$ before $\mathbf{q\rightarrow0}$)}
In the static limit, the factor $S_{m,n}$ can be written as \cite{nanda2023vortical}: 
\begin{equation}
S_{n,n}\left(\mathbf{k},\mathbf{q\rightarrow0},0\right)=\begin{cases}
-\delta\left(\epsilon_{n,\mathbf{k}}\right) & |\nabla_{\mathbf{k}}\epsilon_{n,\mathbf{k}}\cdot\mathbf{q}\tau|\gg1,\\
\frac{1}{\pi}\text{Im}\left[\frac{1}{\epsilon_{n,\mathbf{k}}+\frac{i}{2\tau}}\right] & |\nabla_{\mathbf{k}}\epsilon_{n,\mathbf{k}}\cdot\mathbf{q}\tau|\ll1,
\end{cases}
\end{equation}
\begin{equation}
\begin{split}S_{m,n}\left(\mathbf{k},\mathbf{q\rightarrow0},0\right)\approx\frac{\Theta\left(-\epsilon_{m,\mathbf{k}}\right)-\Theta\left(-\epsilon_{n,\mathbf{k}}\right)}{\epsilon_{m,\mathbf{k}}-\epsilon_{n,\mathbf{k}}} & \  \text{for}\ m \neq n.
\end{split}
\end{equation}
where $\tau^{-1}$ quantifies the strength of disorder. By substituting
this expression for factor $S_{m,n}$ into the Eq. (\ref{eq:33}), and considering
the leading order of $\tau^{-1}$, we obtain
\begin{equation}
\begin{split}
\chi_{ij}^{\text{orb}}(\mathbf{k},\mathbf{q\rightarrow0},0)=\sum_{n}\intop_{\mathbf{k}}\delta\left(\epsilon_{n,\mathbf{k}}\right)m^{\text{orb}}_{i}k_{j}+\sum_{n}\intop_{\mathbf{k}}\Theta\left(-\epsilon_{n,\mathbf{k}}\right)\Omega_{n}^{i} k_{j}\\
\quad \text{for } |\nabla_{\mathbf{k}}\epsilon_{n,\mathbf{k}}\cdot\mathbf{q}\tau|\gg1,
\end{split}
\end{equation}
\begin{equation}
\begin{split}
\chi_{ij}^{\text{orb}}(\mathbf{k},\mathbf{q\rightarrow0},0)=\sum_{n}\intop_{\mathbf{k}}\delta\left(\epsilon_{n,\mathbf{k}}\right)m^{\text{orb}}_{i}k_{j}+\sum_{n}\intop_{\mathbf{k}}\Theta\left(-\epsilon_{n,\mathbf{k}}\right)\Omega_{n}^{i} k_{j}\\
\quad \text{for } |\nabla_{\mathbf{k}}\epsilon_{n,\mathbf{k}}\cdot\mathbf{q}\tau|\ll1.
\end{split}
\end{equation}
where $m^{\text{orb}}_{i}\equiv\frac{i}{2}\left[\left\langle \nabla_{\mathbf{k}} u_{n,\mathbf{k}}\mid\times\left(\epsilon_{n,\mathbf{k}}-H_0(\mathbf{k})\right)\mid\nabla_{\mathbf{k}} u_{n,\mathbf{k}}\right\rangle \right]_{i}$
denotes the $\alpha th$ component of the orbital moment, and $\Omega_{n}^{\alpha}$
is the $\alpha th$ component of the Berry curvature $\mathbf{\Omega}_n\equiv\nabla_{\mathbf{k}}\times \mathbf{K}_{n}$
of $nth$ band.

\subsubsection{Uniform limit ($\mathbf{q\rightarrow0}$ before $\omega\rightarrow0$)}
In the uniform limit, the factor $S_{m,n}$ can be written as:

\begin{equation}
S_{n,n}\left(\mathbf{k},\mathbf{0},\omega\rightarrow0\right)=\begin{cases}
0 & |\omega\tau|\gg1,\\
\frac{1}{\pi}\text{Im}\left[\frac{1}{\epsilon_{n,\mathbf{k}}+\frac{i}{2\tau}}\right] & |\omega\tau|\ll1,
\end{cases}
\end{equation}
\begin{equation}
\begin{split}S_{m,n}\left(\mathbf{k},\mathbf{0},\omega\rightarrow0\right)\approx\frac{\Theta\left(-\epsilon_{m,\mathbf{k}}\right)-\Theta\left(-\epsilon_{n,\mathbf{k}}\right)}{\epsilon_{m,\mathbf{k}}-\epsilon_{n,\mathbf{k}}} & \ \text{for} \ m \neq n.
\end{split}
\end{equation}
the factor $S_{m,n}$ is the same as in the static limit for the disorder
case, however, the intra-band term $S_{n,n}=0$ in the clean case.
Finally, we obtain the susceptibility in uniform limit for clean and
disorder case which is given by:
\begin{equation}
\begin{split}
\chi_{ij}^{\text{orb}}\left(\mathbf{k},\mathbf{0},\omega\rightarrow0\right)=\sum_{n}\intop_{\mathbf{k}}\Theta\left(-\epsilon_{n,\mathbf{k}}\right)\Omega_{n,i}k_{j},
\quad \text{for }  |\omega\tau|\gg1,\\
\end{split}
\end{equation}
\begin{equation}
\begin{split}
\chi_{ij}^{\text{orb}}\left(\mathbf{k},\mathbf{0},\omega\rightarrow0\right)=\sum_{n}\intop_{\mathbf{k}}\delta\left(\epsilon_{n,\mathbf{k}}\right)m^{\text{orb}}_{i}k_{j}+\sum_{n}\intop_{\mathbf{k}}\Theta\left(-\epsilon_{n,\mathbf{k}}\right)\Omega_{n,i}k_{j} \\
\quad \text{for } |\omega\tau|\ll1.
\end{split}
\end{equation}

%\bibliographystyle{unsrt}
%\bibliography{ivelib}
%\end{document}

%\bibliographystyle{plain}
%\bibliographystyle{apsrev4-2}
%\bibliographystyle{unsrt}
%\bibliography{ivelib}

%%%%%%%%%%%%%%

\end{document}